\documentclass[12pt,a4paper]{elsarticle}
\usepackage[utf8]{inputenc}
\usepackage[T1]{fontenc}
\usepackage{amsmath,amsthm}
\usepackage{amsfonts}
\usepackage{amssymb}
\usepackage{caption,subcaption}
\usepackage{graphicx,xcolor}
\usepackage[ruled,vlined]{algorithm2e}
\usepackage[flushleft]{threeparttable}
\usepackage{array,booktabs,makecell}

\newtheorem{theorem}{Theorem}[section]

\newtheorem{remark}[theorem]{Remark}
\usepackage[margin=1in]{geometry}    

\makeatletter
\def\ps@pprintTitle{%
  \let\@oddhead\@empty
  \let\@evenhead\@empty
  \let\@oddfoot\@empty
  \let\@evenfoot\@oddfoot
}
\makeatother

\begin{document}
\begin{frontmatter}  
\title{A DeepParticle method for learning and generating aggregation patterns in multi-dimensional Keller-Segel chemotaxis systems}
	   \author[uoc]{Zhongjian Wang}
        \ead{zhongjian.wang@ntu.edu.sg}
		\author[uci]{Jack Xin}
		\ead{jxin@math.uci.edu}
		\author[hku]{Zhiwen Zhang\corref{cor1}}
		\ead{zhangzw@hku.hk}
		
		\address[uoc]{Division of Mathematical Sciences, School of Physical and Mathematical Sciences, Nanyang Technological University, 21 Nanyang Link, Singapore 637371.}
		\address[uci]{Department of Mathematics, University of California at Irvine, Irvine, CA 92697, USA.}
		\address[hku]{Department of Mathematics, The University of Hong Kong, Pokfulam Road, Hong Kong SAR, China.}		
		\cortext[cor1]{Corresponding author}
\begin{abstract}
We study a regularized interacting particle method for computing aggregation patterns and near singular solutions of a Keller-Segel (KS) chemotaxis system in two and three space dimensions, then further develop the DeepParticle method to learn and generate solutions under variations of physical parameters. The KS solutions are approximated as empirical measures of particles that self-adapt to the high gradient part of solutions. We utilize the expressiveness of deep neural networks (DNNs) to represent the transform of samples from a given initial (source) distribution to a  target distribution at a finite time $T$ prior to blowup without assuming the invertibility of the transforms. In the training stage, we update the network weights by minimizing a discrete 2-Wasserstein distance between the input and target empirical measures. To reduce the computational cost, we develop an iterative divide-and-conquer algorithm to find the optimal transition matrix in the Wasserstein distance. We present numerical results of the DeepParticle framework for successful learning and generation of KS dynamics in the presence of laminar and chaotic flows.  The physical parameter in this work is either the evolution time or the flow amplitude in the advection-dominated regime.  
	\bigskip
	
	
\end{abstract}
\end{frontmatter}
\setcounter{page}{1}

\section{Introduction}

Chemotaxis partial differential equations (PDEs) were introduced by Keller and Segel (KS \cite{keller1970initiation}) to describe the aggregation of the slime mold amoeba Dictyostelium discoideum due to an attractive chemical substance. Related random walk model by Patlak 
was known earlier \cite{Patlak_53}, see \cite{Oth_97} for 
an analysis of basic taxis behaviors (aggregation, blowup, and collapse) based on reinforced random walks. 
Recall a common form of KS model \cite{fatkullin2012study} as follows:
\begin{align}
    \rho_{t} = \nabla \cdot (\mu \, \nabla \rho - \chi\, \rho\,  \nabla c ), \;\; \;\;
    \epsilon \, c_t  = \Delta\, c - k^2 \, c + \rho, \label{KS1}
    \end{align}
    where $\chi, \mu $ ($\epsilon, k$) are positive (non-negative) constants. The model is called elliptic if $\epsilon =0$ (when $c$ evolves rapidly to a local equilibrium), and parabolic if $\epsilon >0$. 
The $\rho$ is the density of active particles (bacteria), and $c$ is the concentration of
chemo-attractant. 
The bacteria diffuse with mobility $\mu$ and drift in the direction of $\nabla c$ with velocity $\chi \nabla c$, where $\chi$ is called chemo-sensitivity.
\medskip

In the simplest regime  $(\epsilon,k)=\mathbf{0}$, the concentration equation 
 becomes the Poisson equation
$  - \Delta c = \rho$. Subject to a suitable boundary condition of $c$, the classical integral representation  
$c=- \mathcal{K}\ast \rho$ holds, 
where $\mathcal{K}$ is the Green's function of the Laplacian, and $\ast$ denotes convolution.
The KS system 
then reduces to a scalar non-local nonlinear advection-diffusion PDE governing the evolution of the density function $\rho$:
\begin{align} 
\rho_t = \mu \, \Delta \, \rho + \chi \, \nabla \cdot \big(\rho \, \nabla( \mathcal{K}\ast \rho)\big). \label{KSequation1}
\end{align}

For modeling chemotaxis in a fluid environment such as ocean  \cite{chemomix_12,chemomix_16,chemomix_yao,chemomix_21,chemomix_22}, people studied equation 
(\ref{KSequation1}) with the advective Lie derivative $\rho$ on the left hand:
\begin{align} 
\rho_t +\nabla \cdot(\rho \, \textbf{v})= \mu \, \Delta \, \rho + \chi \, \nabla \cdot \big(\rho \, \nabla( \mathcal{K}\ast \rho)\big). \label{KSequation2}
\end{align}
The mixing mechanism of the flow field $\mathbf{v}$ is known to slow down or smooth out blowup or aggregation in (\ref{KSequation1}), see analysis in  \cite{chemomix_12,chemomix_16,chemomix_21,chemomix_22} and references therein, and related convection induced smoothing in fluids
\cite{hou2006dynamic,hou2009stabconv}. 
Eq. \eqref{KSequation2} is the macroscopic limit (McKean-Vlasov equation) of the interacting particle system below as $J \uparrow \infty$:
\begin{align}
    d X^{j} = -\,  {\chi \mathcal{M}\over J} \nabla_{X^j} \, 
    \sum_{i=1,\, i\not = j}^{J}\, \mathcal{K} (|X^j - X^i |) \, dt + \textbf{v}(X^j) dt 
+ \sqrt{2\,\mu} \, d W^j, \;\; j=1,\cdots, J,   \label{IPS1}
\end{align}
where $M$ is the conserved total mass (integral of $\rho$), and $W^j$'s are independent Brownian motions in 
$\mathbb{R}^d$. 
\medskip

In this work, $ \textbf{v}$ is a prescribed divergence-free vector field. We shall approximate 
$\rho$ of equation (\ref{KSequation2}) numerically based on the associated interacting particle system in two and three spatial dimensions ($d=2,3$), and carry out a systematic deep learning study (a.k.a. DeepParticle \cite{DP_22}) on the $(\mu, \textbf{v})$ dependency of solutions. 
As we are interested in studying near singular KS solutions, the main challenge for training data collection is to approximate the fields $\rho(\textbf{x})$ and $c(\textbf{x})$ reliably as they intensify. Due to the singular behavior of Green's function, as particles come close to each other, our approach is to regularize $\mathcal{K}$ in  (\ref{IPS1}) for approximating  $\rho(\textbf{x})$ as particles aggregate, in a similar spirit to 
the vortex blob method for fluids (\cite{krasny1991vortex} and references therein) and \cite{blob_aggregation_ck2016}.
\medskip

Deep learning tools have been applied broadly for scientific computing in recent years, such as solving PDEs and their inverse problems; see \cite{Osher_21,DP_22} and references therein.
DeepParticle \cite{DP_22} is based on a particle method for solving a time-dependent physically parameterized PDE, whose solution is approximated by the particle empirical measure (distribution).  A deep neural network (DNN) with physical parameter dependence learns the mapping from the initial particle distribution to 
the particle distribution at time $T$ with training data over sampled physical parameters provided by the particle method. The trained DNN then generates approximate solutions at time $T$ for new physical parameters unseen in the training process. DeepParticle has been successfully designed and trained for learning and generating invariant measures of stochastic interacting particle systems (at $T=\infty$) arising in reaction-diffusion front speeds in three-dimensional chaotic flows \cite{DP_22}. In this paper, we further develop DeepParticle to learn and generate KS solutions exhibiting aggregation behavior at a finite time $T$ before blow-up for a range of diffusivity $\mu$ and advection amplitude values.
\medskip

The rest of the paper is organized as follows. In Section 2, we briefly review the blow-up behavior in the KS model and the regularized particle methods to solve the KS model. In Section 3, we present our DeepParticle method to learn the transport map from an input distribution to a target distribution. Moreover, we will discuss the details of the implementation of our method and how to learn the distributions in particle simulation of the KS model.  In Section 4, we show numerical results to demonstrate the performance of our method, where both 2D and 3D KS chemotaxis systems will be studied. Finally, conclusions and future works will be discussed in section 5.

\section{Regularized interacting particle method of KS model}

\subsection{Blow up behavior in KS model} 
We start from the simplest KS model without advection, namely the Eq. \eqref{KSequation1} with $\mu=\chi=1$ in two spatial dimensions ($d=2$). This system has been extensively studied by many authors; see the survey article \cite{perthame2004pde} and references therein. The conservation of mass holds:
\begin{align}
\frac{d}{ dt}\int_{\mathbb{R}^{2}} \rho(x,t) d x = 0,
\end{align}
which is also true for  Eq. \eqref{KSequation2} when the advection field $\textbf{v}$ is divergence-free.  If we set the total mass $ M:=\int_{\mathbb{R}^{2}} \rho(x,0) d x$, the second moment has a fixed time derivative, i.e.,
\begin{align}\label{second_order_moment}
    \frac{d}{ dt} \int_{\mathbb{R}^{2}}|x|^{2} \rho(x,t)  \, dx=\frac{M}{2 \pi}(8 \pi-M),
\end{align}
where $8 \pi$ is called the critical mass of the KS system. Accordingly, it is well-known that: (1) if $M>8\pi$, the system has no global smooth solutions; (2) if $M=8\pi$, the system has a global smooth solution, which blows up as $t\to\infty$; and (3) if $M<8\pi$, the system has a global smooth solution. 
\medskip

In \cite{chemomix_12,chemomix_16,chemomix_yao}, an extra advection term is introduced to KS-type equations, in order to model organism movement in prescribed fluid flows. Then, the second-moment identity in Eq.\eqref{second_order_moment} and the subsequent blowup vs. global evolution results are no longer valid. By comparison principle, \cite{chemomix_yao} shows that if the total mass is smaller than the critical mass, Eq. (\ref{KSequation2}) has a global smooth solution with smooth initial data. In the cases with supercritical mass, there are only numerical experiments suggesting that the advection, if sufficiently large,  prevents the solutions from blowing up. Later, \cite{chemomix_22} shows that when the flow exerts a  `stretching' effect, the advection field $\textbf{v}$ indeed suppresses the growth or the concentration of chemo-attractant and hence the solution has global regularity and exists for all time.  Examples of stretching flows include hyperbolic flows where $\mathbf{v}(\mathbf{x})=\left( x_{1}, -\frac{1}{d-1}\mathbf{x}_{-}\right)$ and laminar flows where $\mathbf{v}(\mathbf{x})=\left(v\left(\mathbf{x}_{-}\right), \mathbf{0}_{-}\right)$, with $\mathbf{x}_{-}=\left(x_{2}, \ldots, x_{d}\right)$ and $v$ is periodic. However, the case of chaotic flows, or when the amplitude of advection $\textbf{v}$ is not sufficiently large, remains open.

\subsection{Regularized interacting particle methods}

The singular behavior of the governing PDE and the associated Green's function that cause trouble for particle methods is a well-known problem. The vortex blob method \cite{chorin1973discretization} provides a regularization approach to extend vortex sheet motion past the singularity formation time into the physically important roll-up regime, in which the points representing the vortex sheet are replaced by vortices of prescribed and fixed shape. Numerical calculations show regular motion for the centers of the blobs even after the time when a curvature singularity on a vortex sheet is formed. Later, a special form of the vortex blob method \cite{krasny1986desingularization} is used to calculate the roll-up of a periodic vortex sheet resulting from the classical Kelvin-Helmholtz instability. The singular solutions are closely related to the ill-posedness of vortex sheet motion \cite{caflisch1989singular}. Nonetheless, as the regularization parameter approaches zero, the regularized vortex sheet solution converges to a weak solution of the Euler equation \cite{liu1995convergence}. As in vortex blob methods, we formulate a regularized interacting particle method (denoted as IPM from here on) to solve KS chemotaxis systems.
\medskip

 Let us approximate the density function (solution of Eq.\eqref{KSequation1} or Eq.\eqref{KSequation2}) with empirical distributions of particle positions. In SDE \eqref{IPS1} on particle positions, 
the chemo-attractant term $\,  {\chi \mathcal{M}\over J} \nabla_{X^j} \, \sum_{i=1,\, i\not = j}^{J}\, \mathcal{K} (|X^j - X^i |) \, dt$ causes numerical instabilities when particles tend to concentrate. To overcome this difficulty, we replace the singular kernel $\mathcal{K}(\cdot)$ in \eqref{IPS1} by a smoothed approximation $K_{\delta}(\cdot)$, 
such that $K_{\delta}(z)\rightarrow K(z)$ as $\delta \rightarrow 0$, where $\delta>0$ is a regularization parameter. For example, we define 
\begin{align}
    K_{\delta}(z) = K(z)\frac{|z|^2}{|z|^2+\delta^2}. 
\end{align} 
\medskip

Equipped with the kernel $K_{\delta}(\cdot)$, we obtain a system of regularized SDEs for the particles as follows: 
\begin{align}
    d X^{j} = -\,  {\chi \mathcal{M}\over J} \nabla_{X^j} \, 
    \sum_{i=1,\, i\not = j}^{J}\, K_{\delta}(|X^j - X^i |) \, dt+ \textbf{v}(X^j) dt 
+ \sqrt{2\,\mu} \, \, d W^j, \;\; j=1,2,\cdots, J,    
    \label{IPS2}
\end{align}
where $J$ is the number of particles, $X^{j}\in \mathbb{R}^d$ is the position of the $j$-th particle, and $d W^j$'s are mutually independent $d$-dimensional Brownian motions. The convergence of a random particle blob method, similar to (\ref{IPS2}) yet for KS without advection field $\textbf{v}$, is analyzed in  
\cite{rpbm_2017}. For a stochastic interacting particle method using heuristic collision and splitting rules to bypass the singular behavior of Green's function, see \cite{havskovec2009stochastic}.     
\medskip




Representing PDE solutions by particles belongs to the Lagrangian framework.  The Lagrangian methods have several advantages: (1) easy to implement; (2) spatially mesh-free and self-adaptive; and (3)  computational costs scale linearly with the dimension of spatial variables in the underlying stochastic dynamical systems. If we discretize KS system  \eqref{KSequation2} on mesh grids with grid-based methods, e.g. finite element \cite{carrillo2019hybrid} and spectral \cite{shen2020unconditionally} methods, the number of mesh grids depends exponentially on the spatial dimension. The key benefit of the Lagrangian framework in computing KS models is its natural capability to follow the KS solution when a singular behavior is emerging. 

As we shall see, the stochastic particle method based on (\ref{IPS2}) reproduces the well-known aggregation behavior and captures the KS dynamics during the potential blow-up stage of evolution. This is another step forward in our program of computing high gradient solutions in the Lagrangian framework, which has shown encouraging results for a range of multi-dimensional singularly-perturbed advection-diffusion PDEs.
We refer interested readers to our recent progress in developing Lagrangian methods to compute effective diffusivities in chaotic or random flows \cite{WangXinZhang:18,IPM_2020,SharpMMS_21,TDChaoticFlow_22} and KPP front speeds in chaotic flows \cite{IPM_2021}. 
There are also deterministic particle methods (\cite{blob_diff_2019,blob_aggregation_ck2016} and references therein) for a class of KS and degenerate diffusion equations that fall in our DeepParticle framework (see Section 3).
\medskip

Though the IPM in our study here is mesh-free and self-adaptive for solving multi-dimensional KS chemotaxis systems, the computational costs remain high if we want to study the systems under a variation of parameters (e.g. the evolution time $T$ and the amplitude of advection $A$). Also, the number of particles $J$ cannot be too large as the chemo-attractant term has $O(J^2)$ complexity in each time step evolution. The total complexity is then $O(J^2\frac{T}{\Delta t})$, where $\Delta t$ is the time step of discretization for Eq.\eqref{IPS2}. On the other hand, our numerical simulation of Eq.\eqref{IPS2} shows that the distribution at finite time $T$, starting from the same initial distribution, may have continuous dependence on the physical parameters. Therefore in the next section, we will introduce a deep learning algorithm that can learn the continuous dependence of the 
solutions on parameters of the KS models and generate approximated samples with $O(J)$ complexity.



\section{Deep particle method} 
In this section, we introduce a DeepParticle algorithm to learn the features of the transport map from a trivial (input) distribution to a target (output) distribution. The mapping error is measured by the 2-Wasserstein distance.

\subsection{Discrete Wasserstein distance}
\noindent	Given distributions $\mu$ and $\nu$ defined on metric spaces $X$ and $Y$, let us construct a transport map $f^0_{*}:X \to Y$ such that
		$f^{0}_{*}(\mu)= \nu$, where star denotes the push forward of the map. 
For any function $f_*:X \to Y$, the $2$-Wasserstein distance between  $f_*(\mu)$ and $\nu$ is defined by:

\begin{align}\label{def:Wdistance}
	W_{2}(f_*(\mu), \nu):=\left(\inf _{\gamma \in \Gamma(f_*(\mu), \nu)} \int_{Y \times Y} \text{dist}(y', y)^{2} \mathrm{~d} \gamma(y', y)\right)^{1 / 2},
\end{align}
where $\Gamma(f_*(\mu), \nu)$ denotes the collection of all measures on $Y \times Y$ with marginals $f_*(\mu)$ and $\nu$ on the first and second factors respectively, and $\text{dist}(\cdot,\cdot)$ denotes the metric (distance) on $Y$. A straightforward derivation yields:

\begin{align}\label{def:Wdistance1}
    W_{2}(f_*(\mu), \nu)=\left(\inf _{\gamma \in \Gamma(\mu, \nu)} \int_{X \times Y} \text{dist}(f(x), y)^{2} \mathrm{~d} \gamma(x, y)\right)^{1 /2},
\end{align}
where $\Gamma(\mu, \nu)$ denotes the collection of all measures on $X \times Y$ with marginals $\mu$ and $\nu$ on the first and second factors respectively, and still $\text{dist}(\cdot,\cdot)$ denotes the metric (distance) on $Y$.
To design computational methods, we approximate distributions $\mu$ and $\nu$ by empirical distribution functions:
$	\mu=\frac{1}{N}\sum_{i=1}^{N}\delta_{x_i}$ and  $	\nu=\frac{1}{N}\sum_{j=1}^{N}\delta_{y_j}$, where $N$ is the number of samples in the empirical distributions. Under the setting of learning distribution from interacting particle methods, we take $N<J$ sub-samples from the terminal time position of system \eqref{IPS1} to represent the distribution. We will re-sample them every $1000$ steps of training. The preceding technique is usually referred to as \emph{mini-batch} in the deep learning literature.
\medskip

Any joint distribution in $\Gamma(\mu,\nu)$ can be approximated by an $N\times N$ doubly stochastic matrix \cite{sinkhorn1964relationship}, denoted as transition matrix,  $\gamma=(\gamma_{ij})_{i,j}$ satisfying:   
	
	\begin{align}
	\label{def:bistochasticity}
		\gamma_{ij} \geq 0; \quad \quad
	 \forall j, ~\sum_{i=1}^{N}\gamma_{ij}=1;
	 \quad \quad  
	 \forall i, ~\sum_{j=1}^{N}\gamma_{ij}=1.
	\end{align}	

Then \eqref{def:Wdistance1} becomes
\begin{align}\label{def:Wdistance_approx}
\hat{W}(f):=\left(\inf_{\gamma \in \Gamma^N}\frac{1}{N}\sum_{i,j=1}^N\, \text{dist}(f(x_i),y_j)^2 \gamma_{ij}\right)^{1/2}.
\end{align}
$\hat{W}(f)$ in \eqref{def:Wdistance_approx} has a simple intuitive interpretation: given a $\gamma \in \Gamma(\mu, \nu)$ and any pair of locations $(x,y)$, the value of $\gamma(x,y)$ tells us what proportion of $f_*(\mu)$ mass at $f(x)$ should be transferred to $y$, in order to reconfigure $f_*(\mu)$ into $\nu$. Computing the effort of moving a unit of mass from $f(x)$ to $y$ by $\text{dist}(f(x),y)^{2}$ yields the interpretation of $\hat{W}$ as the minimal effort (optimal transportation \cite{villani2021topics}) to reconfigure $f_*(\mu)$ mass distribution into that of $\nu$. 

\subsection{Training data and neural network configuration}
Note that given any fixed set of $\{x_i\}_{i=1}^N\subset \mathbb{R}^d$ and $\{y_j\}_{j=1}^N\subset \mathbb{R}^d$ (training data), we have derived Eq.\eqref{def:Wdistance_approx} to be minimized by gradient descent. In addition, we aim to find a network that can represent the change of target distribution over some physical parameters. In such a scenario, more than one set of data ($\{x_i\}$ and $\{y_j\}$ consists of one set of data) should be assimilated. More precisely, let the total number of data set be denoted as $N_{dict}$. Then we have $N_{dict}$ pairs of i.i.d. samples of input and output distribution, denoted by $\{x_{i,r}\}$ and $\{y_{j,r}\}$ for $r=1\cdots N_{dict}$. Associated with the $r$-th data set ($\{x_{i,r}\}$ and $\{y_{j,r}\}$), we assume that there is a physical parameter $\eta_r\in \mathbb{R}^p$. To represent this in the network, we encode $\eta_r$ to each data in the set, i.e., the input of the network is $\{(x_{i,r},\eta_r)\}_i \subset \mathbb{R}^{d+p}$. This procedure is also called \textit{padding} in the literature.  The output of the network is then denoted by $f_\theta(x;\eta)$ where $\theta$'s are all trainable parameters of the network, i.e., the weights in the neural network. 
  
We want to emphasize that our neural network is very compact. Between the input layer ($l_0$) and output layer ($l_6$), there are 5 latent layers, where each layer is $30$ in width. The adjacent layers $l_i$ and $l_{i+1}$ are fully connected, i.e., 
 \begin{align}
     l_{i+1}=\tanh(W_i(l_{i})+b_i) \quad i=0,\cdots 4,
 \end{align}
 where $W_i$ is (weight) matrix width of $l_{i+1}$ and $l_{i}$, $b_i$ is the (bias) vector with the same dimension as $l_{i}$, and  $\tanh(\cdot)$ is the activation function. For the output layer, the formula is similar except we do not apply the activation function.  In the case of $d=2$ and $p=1$ (e.g. the first numerical example in computing blowup behavior of KS without advection in Section \ref{sec:eg1}), there are $3902$ parameters (weight and bias) to learn. In 3D cases, we find that the network performs well even without altering the width of latent layers, where the parameter number is $3963$; see Table \ref{tab:parameterno} for the detailed shape and size.
  \begin{table}[h]
  \centering 
  \begin{threeparttable}
    \begin{tabular}{ccr}
    \toprule
Parameter & Shape & Size \\ 
\midrule
$W_0$ & $30\times 4$ & 120  \\
$b_0$ & $30\times 1$ & 30  \\
$W_{1:4}$  & \hspace{0.5em}$30\times 30$ & $900 \times 4$ \\
$b_{1:4}$ & $30\times 1$ & $30\times 4$  \\
$W_5$ & \hspace{1em}$3\times 30$  & 90  \\
$b_5$ & \hspace{0.5em}$3\times 1$ & 3  \\
Total & & 3963\\
  \bottomrule
    \end{tabular}
\end{threeparttable}
 \caption{Shape and size of network parameters to learn in case of $d=3$ and $p=1$}
 \label{tab:parameterno}
\end{table}
 \begin{remark}
 Compared with the par-net approach developed in \cite{DP_22}, we found that  \textit{padding} achieves similar performance when learning distribution in particle simulation of the KS model. 
 \end{remark}
 
  In our computation, by equipping $\mathbb{R}^d$ with Euclidean metric, we can get the training loss function as follows:
\begin{align}\label{def:penalty-1}
	\hat{W}^2(f_\theta):=\frac{1}{Nn_\eta}\sum_{r=1}^{n_{\eta}}\left(\inf_{\gamma_r \in \Gamma^N}\sum_{i,j=1}^N|f_\theta(x_{i,r};\eta_r)-y_{j,r}|^2\gamma_{ij,r}\right).
\end{align}

The goal of DeepParticle algorithm is then to find $(\theta,\{\gamma_r\})$ through minimizing
\begin{align}\label{def:goal}
	P(\theta,\{\gamma_r\}):=\sum_{r=1}^{n_{\eta}}\sum_{i,j=1}^N\left(|f_\theta(x_{i,r};\eta_r)-y_{j,r}|^2\gamma_{ij,r}\right).
\end{align}

\subsection{Iterative method in finding transition matrix $\gamma$}\label{sec:iterative_method}
Notice that, with fixed $\theta$, finding the transition matrix $\gamma$ to calculate discrete Wasserstein distance in \eqref{def:penalty-1} is a linear programming problem with a degree of freedom $N^2$. When the number of particles $N$ becomes large, it is expensive to go by a conventional algorithm, e.g. interior point algorithm or simplex algorithm.  We now present a mini-batch linear programming algorithm to find the best $\gamma$ for each  
 inner sum of \eqref{def:penalty-1}, while suppressing $\eta_r$ dependence in $f_\theta$ for notational simplicity. 

\medskip
 
The problem \eqref{def:penalty-1} is a linear programming problem on the bounded convex set $\Gamma^N$ of vector space of real $N\times N$ matrices. By Choquet's theorem, this problem admits solutions that are extremal points of $\Gamma^N$. The set of all doubly stochastic matrix $\Gamma^N$ is viewed as Birkhoff polytope. The Birkhoff–von Neumann theorem \cite{schrijver2003combinatorial} states that such polytope is the convex hull of all permutation matrices, i.e., those matrices such that $\gamma_{ij}=\delta_{j,\pi(i)}$ for some permutation $\pi$ of $\{1,...,N\}$, where $\delta_{jk}$ is the Kronecker symbol. 
 \medskip
 
 Our algorithm is defined iteratively. We start from a permutation matrix, e.g., $\gamma_{ij}=\delta_{ij}$. In each iteration, we randomly select columns and corresponding rows such that the submatrix is a permutation matrix. Then, the entries of the submatrix consist of  a linear programming problem under the constraint that maintains column-wise and row-wise sums equal to one. To be precise, we randomly choose  $\{i_k\}_{k=1}^M$, ($M\ll  N$)  from $\{1,2, \cdots, N\}$ without replacement. Then, 
 we select the indices $j_k$'s such that $\gamma_{i_k,j_k}=1$.  The cost function of the sub-problem is 
 \begin{align}\label{def:lp-sub-cost}
	C(\gamma^*):=\sum_{k,l=1}^M|f_\theta(x_{i_k})-y_{j_l}|^2\gamma^*_{i_kj_l}
\end{align}
subject to
\begin{align}\label{def:lp-sub-cons}
\left\{	\begin{array}{l}
		\sum_{k=1}^M \gamma^*_{i_k,j_l}=1\quad\forall l=1,\cdots,M\\
			\sum_{l=1}^M \gamma^*_{i_k,j_l}=1\quad\forall \, k=1,\cdots,M\\
	\gamma^*_{i_kj_l}\geq 0 \quad \forall \, k,l = 1,\cdots,M.
\end{array}\right.
\end{align}
Then, the minimizer $\gamma^*$ is again a permutation matrix.  The linear programming sub-problem of much smaller size is solved by the interior point method  \cite{wright1997primal}. In addition, as the goal is to find a permutation matrix, we set the tolerance of the interior point to be relatively large and project the resulting approximation to a permutation matrix. In our approach, we terminate the iteration of the sub-problem until $\min_k(\max_l \gamma_{i_k,j_l}^*)>0.5$. This is to ensure that there is a unique large-value entry in each column. As a projection, we update $\gamma$ by, 
\begin{align}
    \gamma_{ij}=\left\{ \begin{array}{l}
         1 \quad \text{if}\ j=\arg\max_l \gamma^*_{il}, \\
         0 \quad \text{otherwise.}
    \end{array}\right.
\end{align}

We observe that the global minimizer of $\gamma$ in  \eqref{def:penalty-1} is also the solution of sub-problems \eqref{def:lp-sub-cost} with arbitrarily selected rows and columns, subject to the row and column partial sum values of the global minimum. The selection of rows can be one's own choice. In our approach, in each step after gradient descent, we randomly select rows to solve the optimization problem iteratively.
\medskip

Compared with the Wasserstein Generative Adversarial Network (WGAN) proposed in \cite{arjovsky2017wasserstein}, \eqref{def:goal} is a Min-Min optimization problem. Both Adam gradient descent of $\theta$ and mimi-batch optimization of $\gamma$ are iteratively defined. We then alternatively update $\theta$ and $\gamma$ to seek a global minimizer of \eqref{def:goal}.
\medskip

The computational cost of finding optimal $\gamma$ increases as $N$ increases, however, the network itself is independent of $\gamma$. After training, our network acts as a sampler for some target distribution $\nu$ 
without the assumption that $\nu$ should have a closed-form distribution function. At this stage, the input data is no longer limited by training data, and an arbitrarily large amount of samples approximately obeying $\nu$ can be generated through $\mu$ (uniform distribution).


\subsection{Full Training Algorithm}
The full training process is outlined in Alg. \ref{alg:1}, and carried out on a quad-core CPU desktop with an RTX2070 8GB GPU at UC Irvine. The training data is collected from the first-order explicit IPM that solves the regularized SDE system (\ref{IPS2}) by Euler's method in time. The IPM is also the reference solver for evaluating DeepParticle generations. 

\begin{algorithm}[htbp]
	\SetAlgoLined
		Randomly initialize weight parameters  $\theta$ in network $f_\theta:\mathbb{R}^d\to \mathbb{R}^d$; \\
	\Repeat{given number of training mini-batches, $N_{dict}$}{
	\For{ physical parameter set $r\gets0$ \KwTo $n_{\eta}$ }{
	randomly select $\{x_{i,r}\}$, $\{y_{j,r}\}$, $i,j=1:N$ from i.i.d. samples of input and target distribution (generated by the IPM) with respect to physical parameter $\eta_r$;\\
	$\gamma_{ij,r}=\delta_{i,j}$, i.e., initialize it as a permutation matrix;\\
	}
	\If{not the first training mini-batch}{
		\For{ physical parameter set $r\gets0$ \KwTo $n_{\eta}$ }{
		$\delta P_r=+\infty$\\
			\While{$|\delta P_r|<tol$}{
		$P_r=\sum_{i,j=1}^N|f_\theta(x_{i,r},\eta_r)-y_{j,r}|^2\gamma_{ij,r}$;\\
		randomly choose $\{i_{k,r}\}_{k=1}^M$ from $\{1,2, \cdots, N\}$ without replacement; \\
		find $\{j_{k,r}\}_{k=1}^M$ such that $\gamma_{i_{k,r}j_{k,r},r}=1$;\\
		solve the linear programming sub-problem \eqref{def:lp-sub-cost}-\eqref{def:lp-sub-cons} 
		and get $\gamma_r^*$ that is another permutation matrix;\\
		update $\{\gamma_{i_{k,r}j_{l,r}}\}_{k,l=1}^M$ with $\{\gamma^*_{i_{k,r}j_{l,r}}\}_{k,l=1}^M$;\\
		$\delta P_r=\sum_{i,j=1}^N|f_\theta(x_{i,r},\eta_r)-y_{j,r}|^2\gamma_{ij,r}^*-P_r.$
	}
	}
	}
	\Repeat{given steps for each training mini-batch}{
		$P=\sum_{r=1}^{N_r}\sum_{i,j=1}^N|f_\theta(x_{i,r},\eta_r)-y_{j,r}|^2\gamma_{ij,r}$;\\
		$\theta\leftarrow \theta - \delta_1\nabla_{\theta} P$, $\delta_1$ is the learning rate;\\
		\Repeat{given linear programming steps, $N_{LP}$}{
					\For{ physical parameter set $r\gets0$ \KwTo $n_{\eta}$ }{
		$P_r=\sum_{i,j=1}^N|f_\theta(x_{i,r},\eta_r)-y_{j,r}|^2\gamma_{ij,r}$;\\
		randomly choose $\{i_{k,r}\}_{k=1}^M$, from $\{1,2, \cdots, N\}$ without replacement; \\
		find $\{j_{k,r}\}_{k=1}^M$, such that $\gamma_{i_{k,r}j_{k,r},r}=1$;\\
		solve the linear programming sub-problem \eqref{def:lp-sub-cost}-\eqref{def:lp-sub-cons} 
		and get $\gamma_r^*$ which is another permutation matrix;\\
		update $\{\gamma_{i_{k,r}j_{l,r}}\}_{k,l=1}^M$ with $\{\gamma^*_{i_{k,r}j_{l,r}}\}_{k,l=1}^M$.\\
	}
	}
	}
	} 
	$\mathbf{Return}$
	\caption{DeepParticle Learning}\label{alg:1}
\end{algorithm}
\section{Numerical Examples}
\subsection{2D KS Simulation and Generation in the Absence of Advection}\label{sec:eg1}
First, we consider the KS model without advection, namely the Eq.\eqref{KSequation1}. In addition, we choose $\mu=\chi=1$. A straightforward derivation shows that if the initial mass $M>8\pi$ and has a finite second moment, the system will blow up in finite time. 
\medskip

\begin{figure}[htbp]
    \centering
     \begin{subfigure}{0.31\textwidth}
    \includegraphics[width=\linewidth]{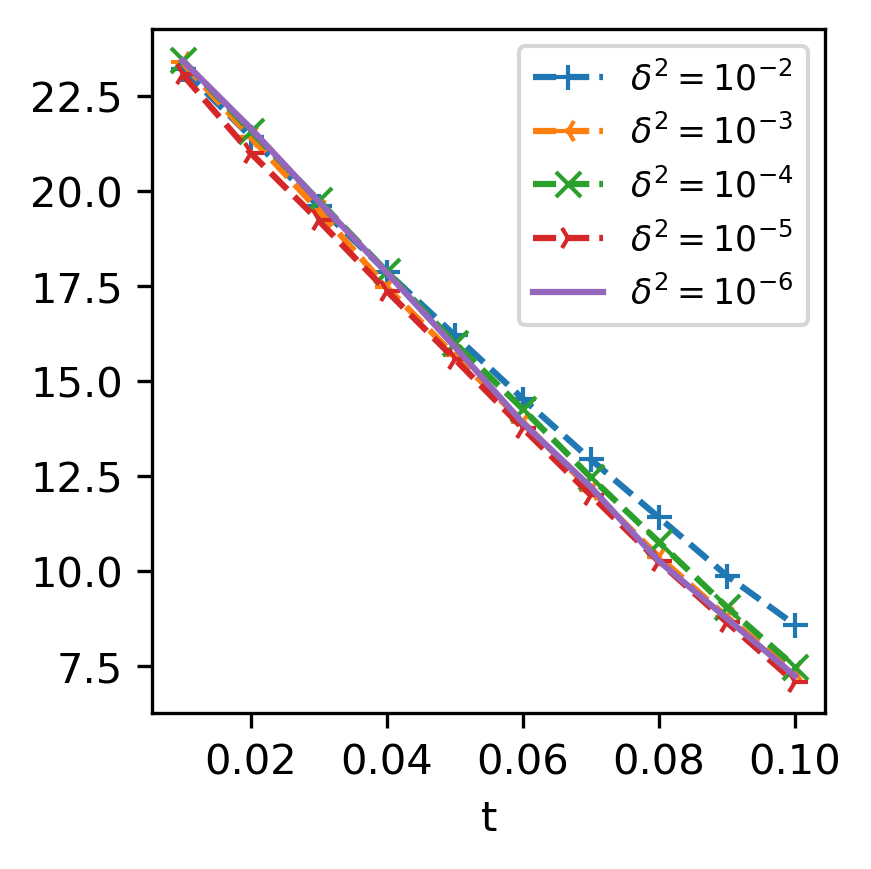}
    \caption{Second moment from IPM with various regularization parameters}
    \end{subfigure}
     \begin{subfigure}{0.31\textwidth}
    \includegraphics[width=\linewidth]{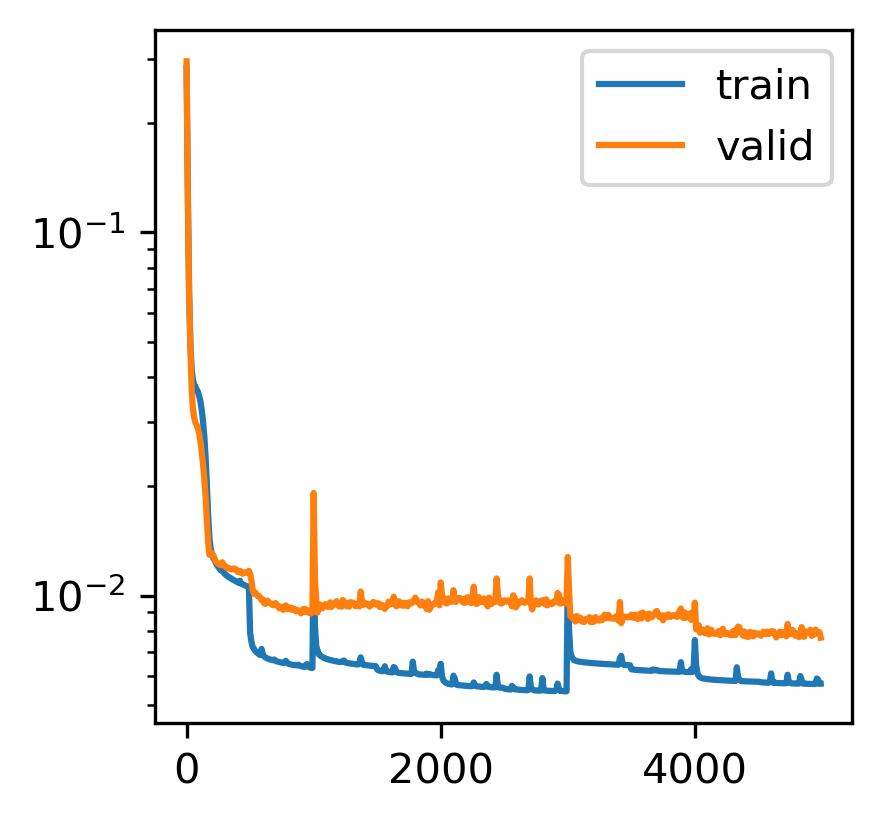}
    \caption{Training and validation loss of $\hat{W}^2$}
    \end{subfigure}
     \begin{subfigure}{0.31\textwidth}
   \includegraphics[width=\linewidth]{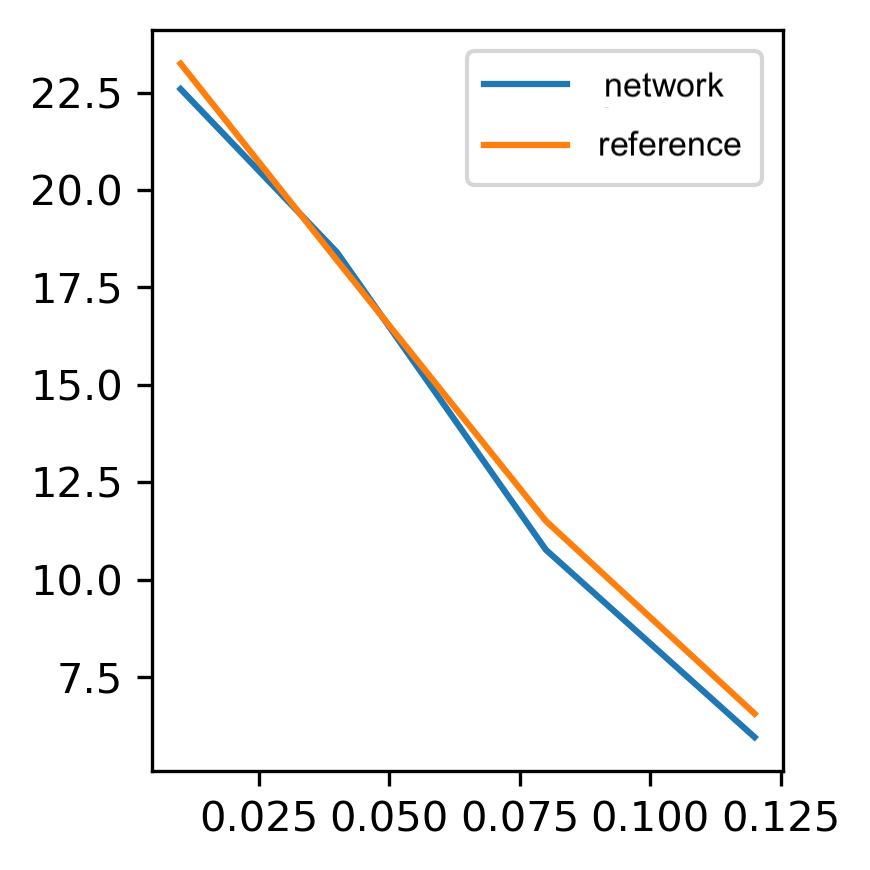}
    \caption{Comparison of the second moment}
    \end{subfigure}
    \caption{Performance of the IPM (reference) and DeepParticle (network) algorithms.}
    \label{fig:eg1_training}
\end{figure}

As the first numerical example, we consider learning the change of distribution depending on evolution time $T$ starting from the initial distribution. The initial distribution is assumed to be a uniform distribution on a ball with a radius $1$ centered at the origin. Assuming the total mass is $16 \pi$, we have then the initial second moment as $8\pi$, which is the same as in \cite{chemomix_yao}. By Eq.\eqref{second_order_moment}, the system will blow up when $t>0.125$. Before applying the deep learning algorithm, we first investigate the error of regularized method. In Fig.\ref{fig:eg1_training}(a), we show the second moment of IPM simulation with various regularization parameters. We see that except for the value  $\delta^2=10^{-2}$, a smaller regularization parameter $\delta$ does not affect the accuracy of the IPM simulation when $t\in[0,0.1]$.
\medskip

Then, we turn to investigate the performance of the proposed DeepParticle algorithms. To get the training data, we apply the IPM with regularization parameter $\delta^2=10^{-3}$ for $T=0.1$ with $J=10000$ particles. We keep the snapshots of the empirical distribution every $0.05$ time interval. During the training process, we consider a mini-batch of size $8\times 2000$, which means that we take $N_{dict}=8$ sets of training data at various times $t$. Namely, we replace the general notation for physical parameter $\eta$ in the network expression $f_\theta(x,\eta)$. Notice that the physical parameter $\eta$ is the evolution time $t$ in this example. In each data set (mini-batch), we have $N=2000$ subsamples from $J=10000$ samples from the IPM. We apply the Adam stochastic gradient descent method to learn the parameters (weights and bias) in the network. We renew the mini-batch every $1000$ steps and renew $\gamma$ every $200$ steps. 

In Fig.\ref{fig:eg1_training}(b), we show the training loss of $\hat{W}^2(f_\theta)$ computed by Eq.\eqref{def:penalty-1}. Since we do not renew $\gamma$ at every step of the gradient descent of parameters, the training loss is not uniformly decreasing. However, the loss is on a decreasing trend overall, which shows that the network successively learns the feature of the distribution as training progresses. In addition, to validate the generalization capabilities of the model, we generate a set of validation distributions at time $t=[0.005,0,025,0,045,0.065,0.095]$ and compute the Wasserstein distance between output distribution and validation distribution every $10$ steps. The trend is similar to the training loss. 
\medskip

After the training process, we denote the trained parameters (weights and bias) in the network as $\theta_1$ and evaluate the network $f_{\theta_1}(x,t)$ at various time $t$ with a larger batch $\{x_i\}_{i=1}^{J'}$ of size $J'=1M$. Though $J'\gg J=10K$, the computational cost of the network evaluation, $O(J')$, is obviously smaller than that of the IPM, $O(J^2 \frac{t}{\Delta t})$ as discussed in Section 2.2. In this example, the DeepParticle method takes $0.015$ seconds to generate $J'=1M$ samples while the reference IPM needs $120$ seconds to generate $10K$ samples.   

We emphasize that the generated output of the network is the image of a continuous map $f_{\theta_1}$ acting on finer samples of the input distribution, which provides an efficient method to generate samples for the output distribution. In fact, it is not a direct replica of training data when the physical parameter (e.g. the time $t$) overlaps with a value in the training set. The generated output of the network will not congregate on points in the training empirical distribution; see Fig.\ref{fig:Kflow}. Leveraging this feature, we can further use these generated samples as a warm-up step to accelerate the IPM simulation. A similar idea appears in our recent paper \cite{DP_22} when computing invariant measures.

In Fig.\ref{fig:eg1_training}(c), we plot the second moment of the particles obtained by the IPM solver (denoted as reference solution) and by our DeepParticle (denoted as network) generation. We see that the slope of the second moment by reference IPM solver deviates from the network output after $t=0.075$. This is due to the fact that the regularization parameter $\delta >0$ in the IPM. For training data, we rely on the IPM solver most of the time when it has good accuracy prior to the time when the $\delta$ effect kicks in.

\begin{figure}[htbp]
    \centering
    \begin{subfigure}{0.49\textwidth}
    \includegraphics[width=\linewidth]{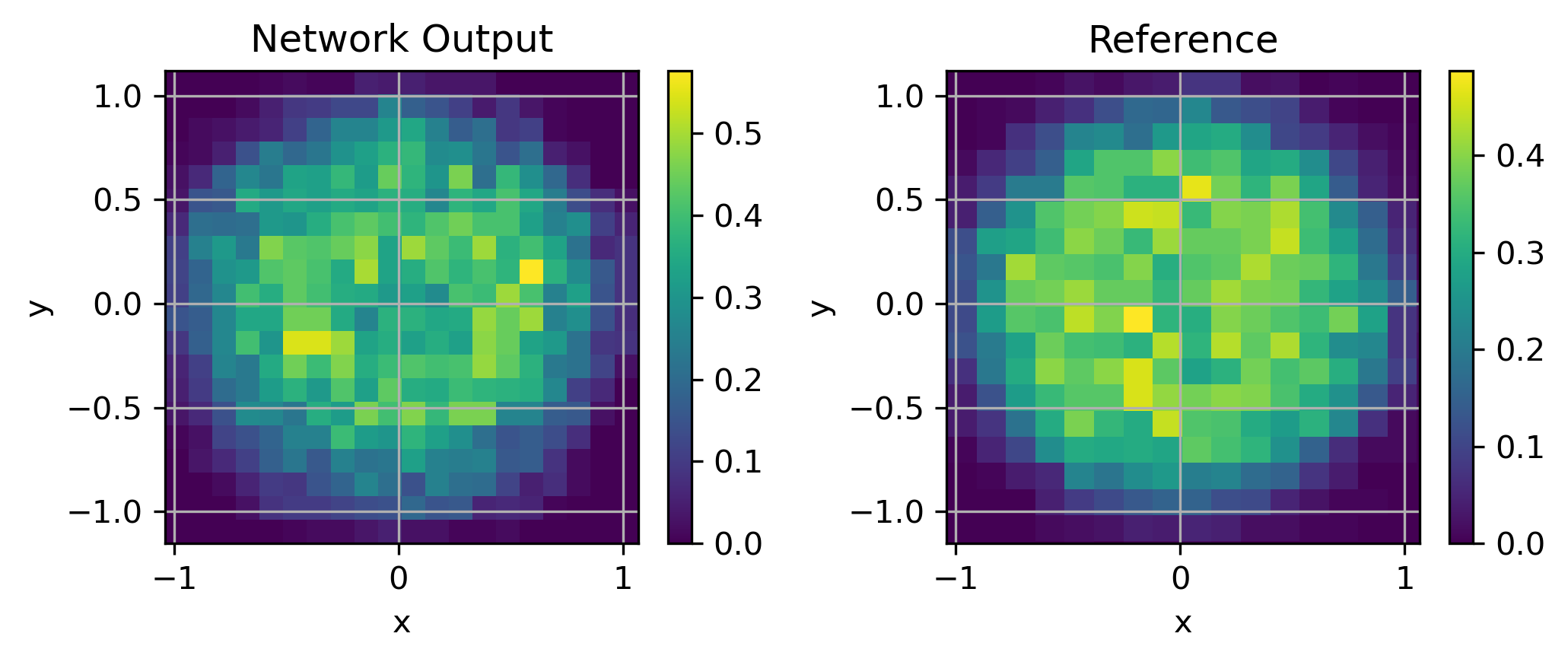}
    \caption{$t=0.01$}
    \end{subfigure}
    \begin{subfigure}{0.49\textwidth}
    \includegraphics[width=\linewidth]{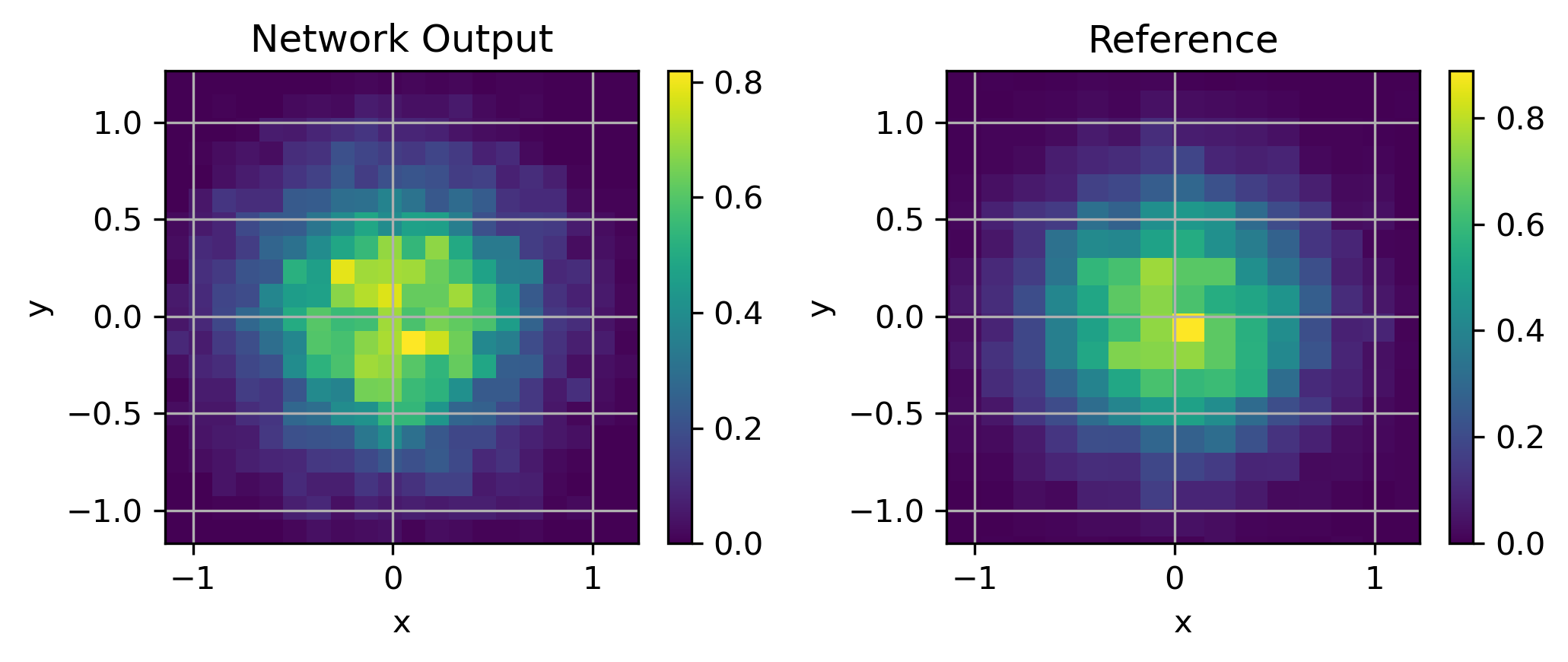}
    \caption{$t=0.04$}
    \end{subfigure}\\
    \begin{subfigure}{0.49\textwidth}
    \includegraphics[width=\linewidth]{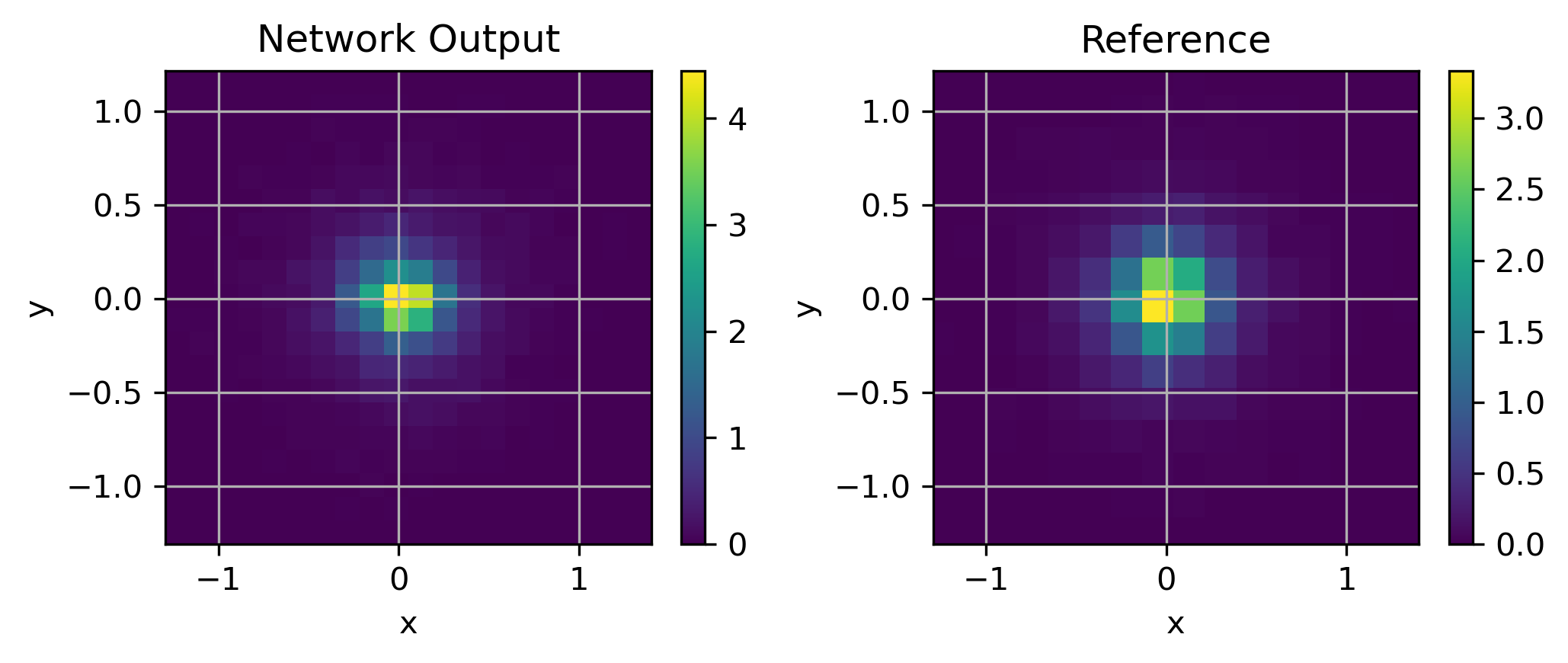}
    \caption{$t=0.1$ (the largest $t$ in the training data)}
    \end{subfigure}
    \begin{subfigure}{0.49\textwidth}
    \includegraphics[width=\linewidth]{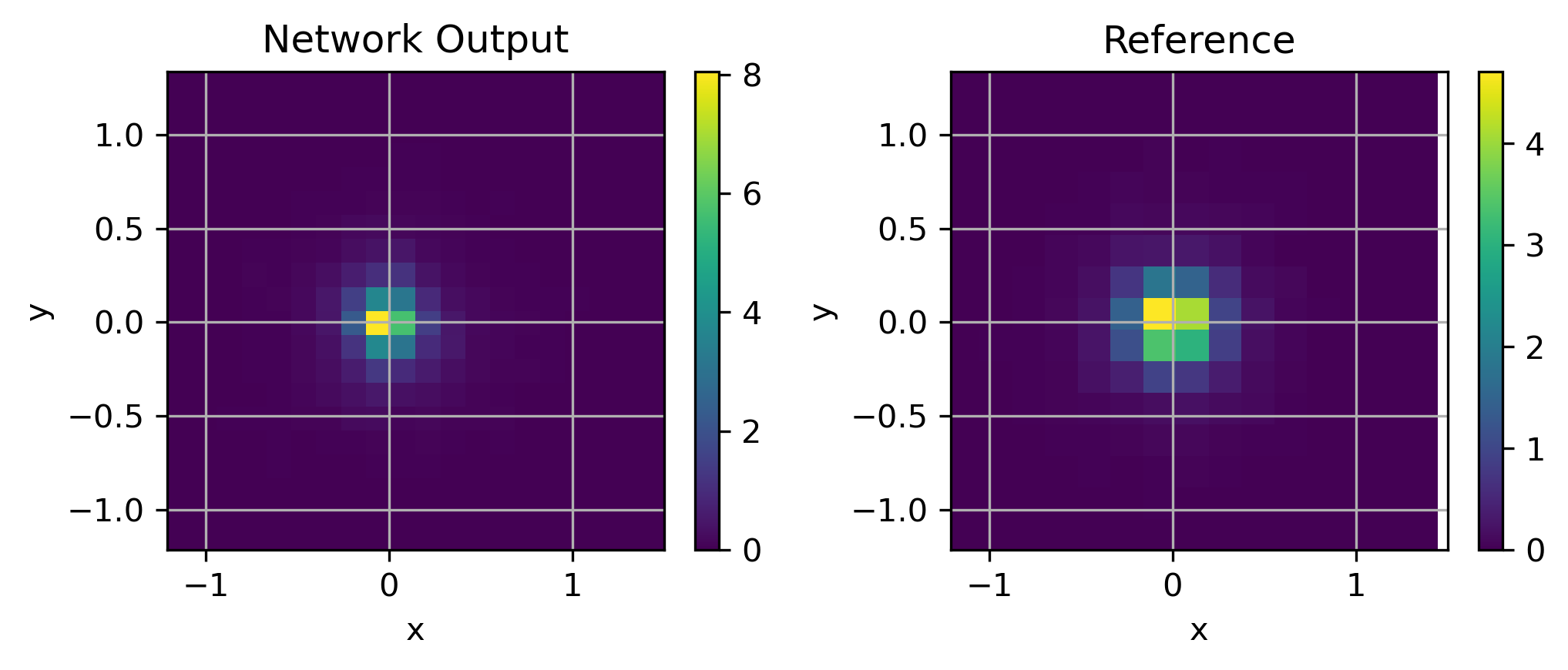}
    \caption{$t=0.12$ (near blowup time)}
    \end{subfigure}
    \caption{Comparison of particle distributions obtained by the DeepParticle method (network output) and IPM (reference) solver at different times and $A=0$.}
    \label{fig:eg1_samples}
\end{figure}

In Fig.\ref{fig:eg1_samples}, we plot the histogram of the network output (using $10^6$ particles) and reference distribution (using $10^4$ particles) at different times $t$. We see that the network learns the feature of particle concentration and even has a more concentrated output than the reference solver near the blowup time. 
\subsection{KS Simulation and Generation in the Presence of 2D Laminar Flows}
Next, we consider the KS model with advection, i.e., $\textbf{v}\not=0$ in the Eq.\eqref{KSequation2}. 
In this case, the movement of the organism is not only driven by the chemotaxis gradient but also driven by some given environmental fluid velocity field. In \cite{chemomix_yao}, the blow-up behavior of the KS model given various strengths of environment velocity is investigated. Now we let
\begin{align}\label{2Dlaminer}
    \textbf{v}(x,y)= A \, \begin{pmatrix}\exp(-y^2)\\0\end{pmatrix},
\end{align}
which represents a laminar flow of amplitude $A$ traveling along $x$ direction with $y$-dependent speed. There are two physical parameters to learn in the model: the amplitude $A$ of the advection field and the evolution time $t$. 

\begin{figure}[htbp]
    \centering
    \begin{subfigure}{0.48\textwidth}
    \includegraphics[width=\linewidth]{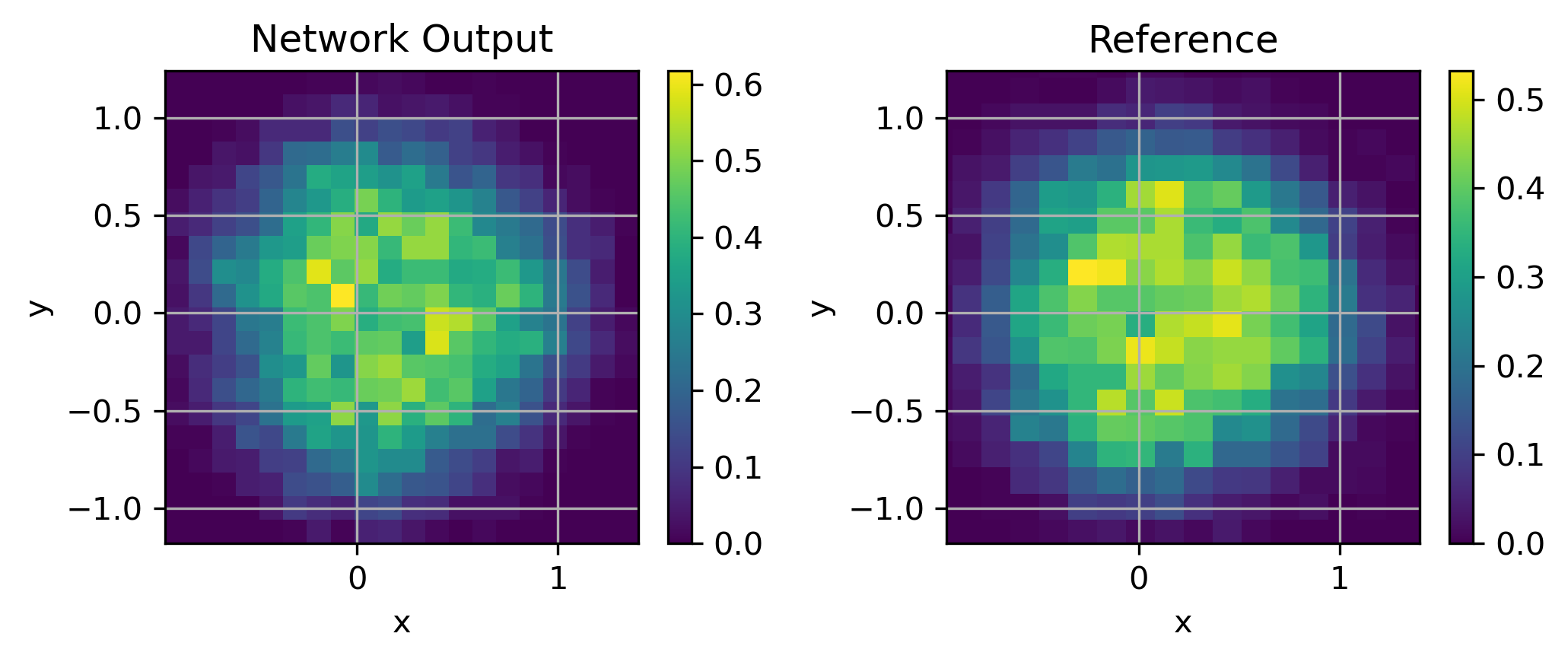}
    \caption{$A=10$}
    \end{subfigure}
    \begin{subfigure}{0.48\textwidth}
    \includegraphics[width=\linewidth]{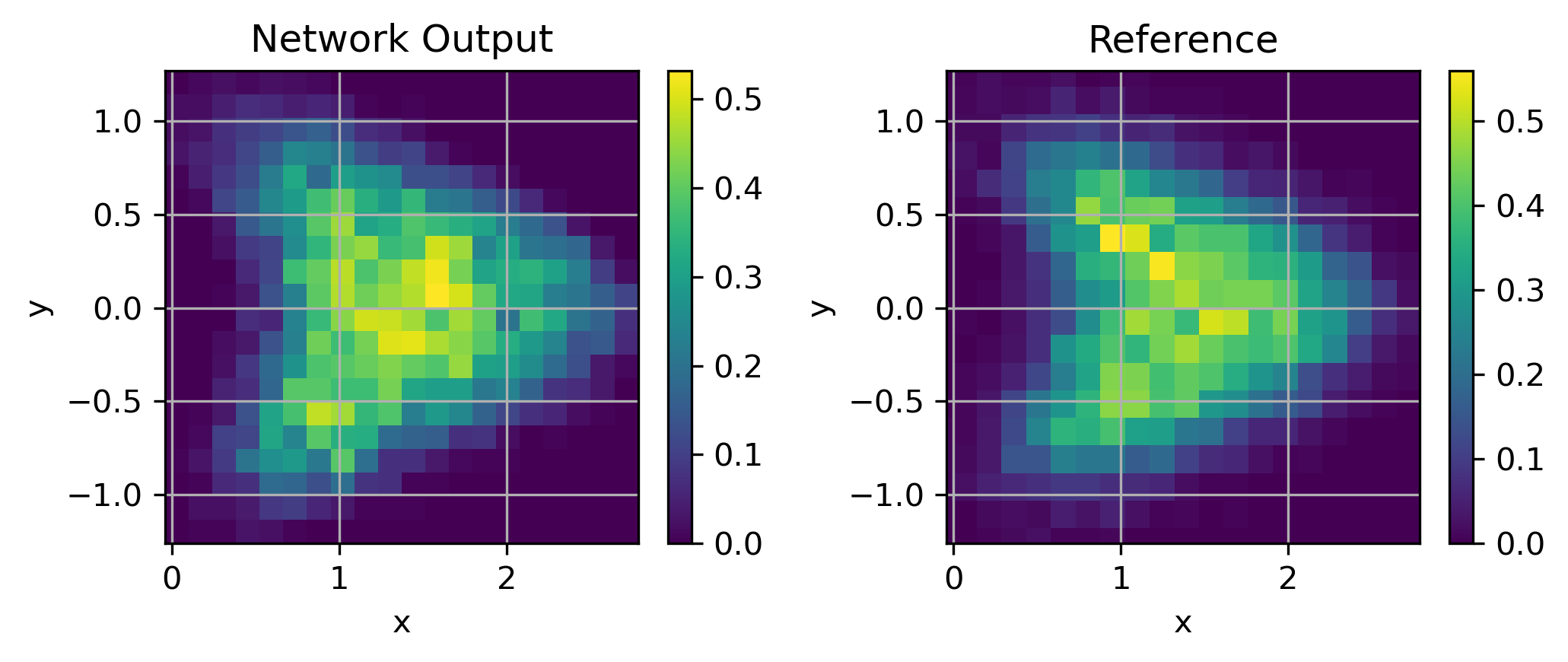}
    \caption{$A=80$ (interpolation)}
    \end{subfigure}\\
    \begin{subfigure}{0.48\textwidth}
    \includegraphics[width=\linewidth]{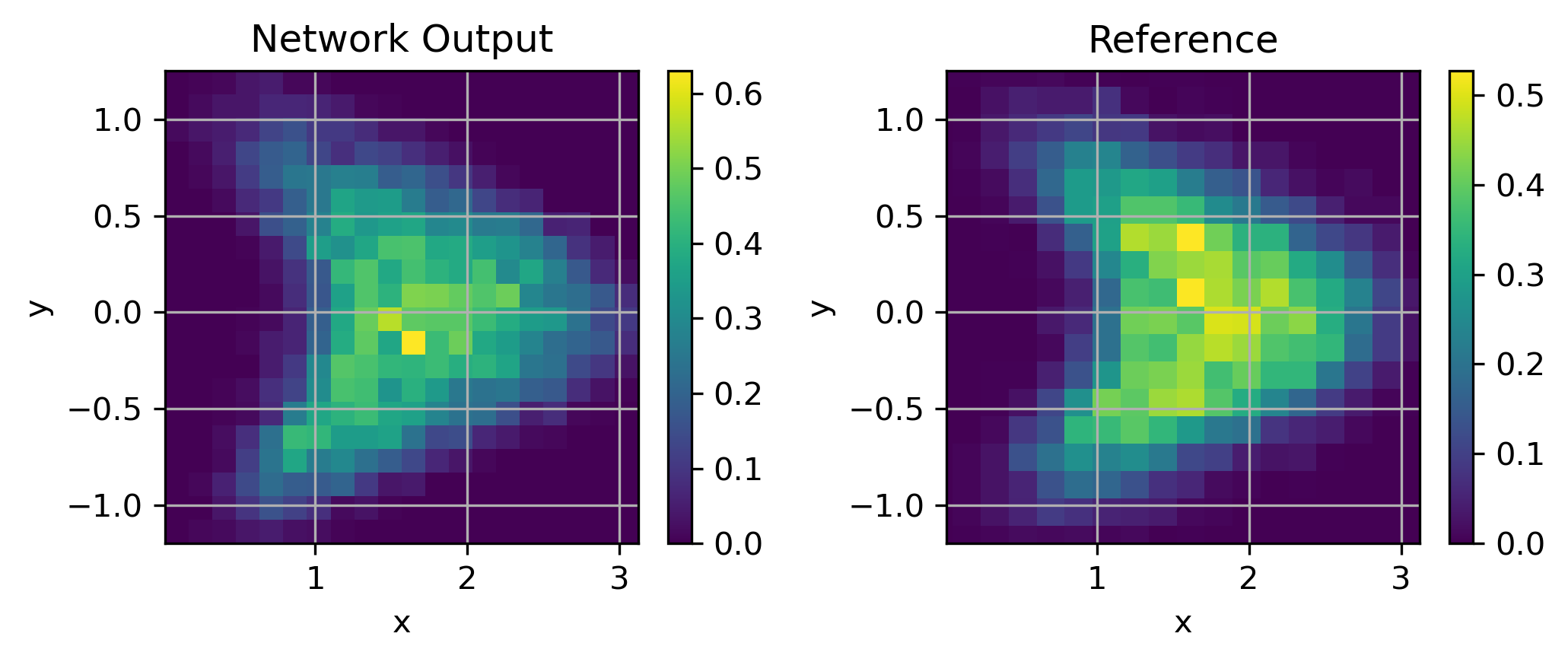}
    \caption{$A=100$ (the largest $A$ in the training set)}
    \end{subfigure}
    \begin{subfigure}{0.48\textwidth}
    \includegraphics[width=\linewidth]{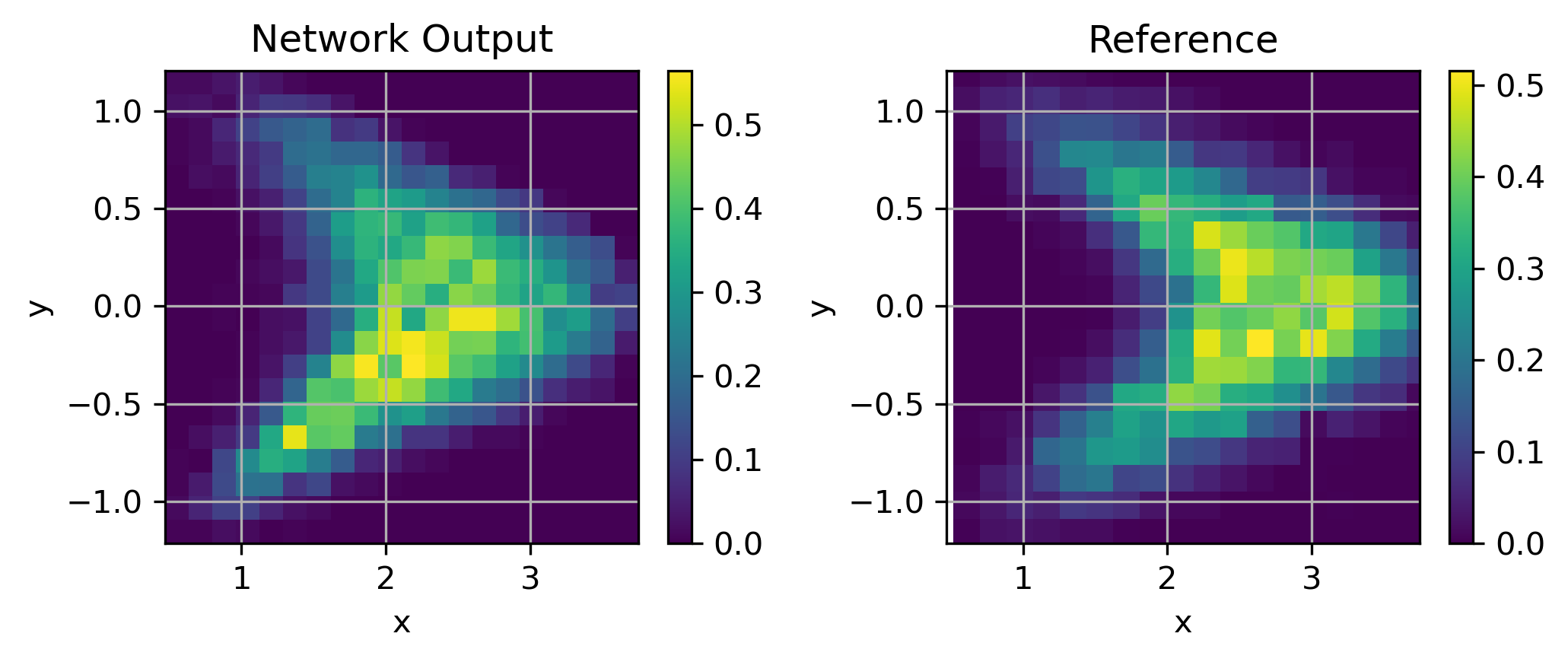}
    \caption{$A=150$ (extrapolation)}
    \end{subfigure}
    \caption{Learning particle aggregation at different $A$ values with $t=0.02$ fixed. Both the interpolation and extrapolation performances of the network are tested.}
    \label{fig:2DlaminerC}
\end{figure}

\begin{figure}[htbp]
    \centering
    \begin{subfigure}{0.48\textwidth}
    \includegraphics[width=\linewidth]{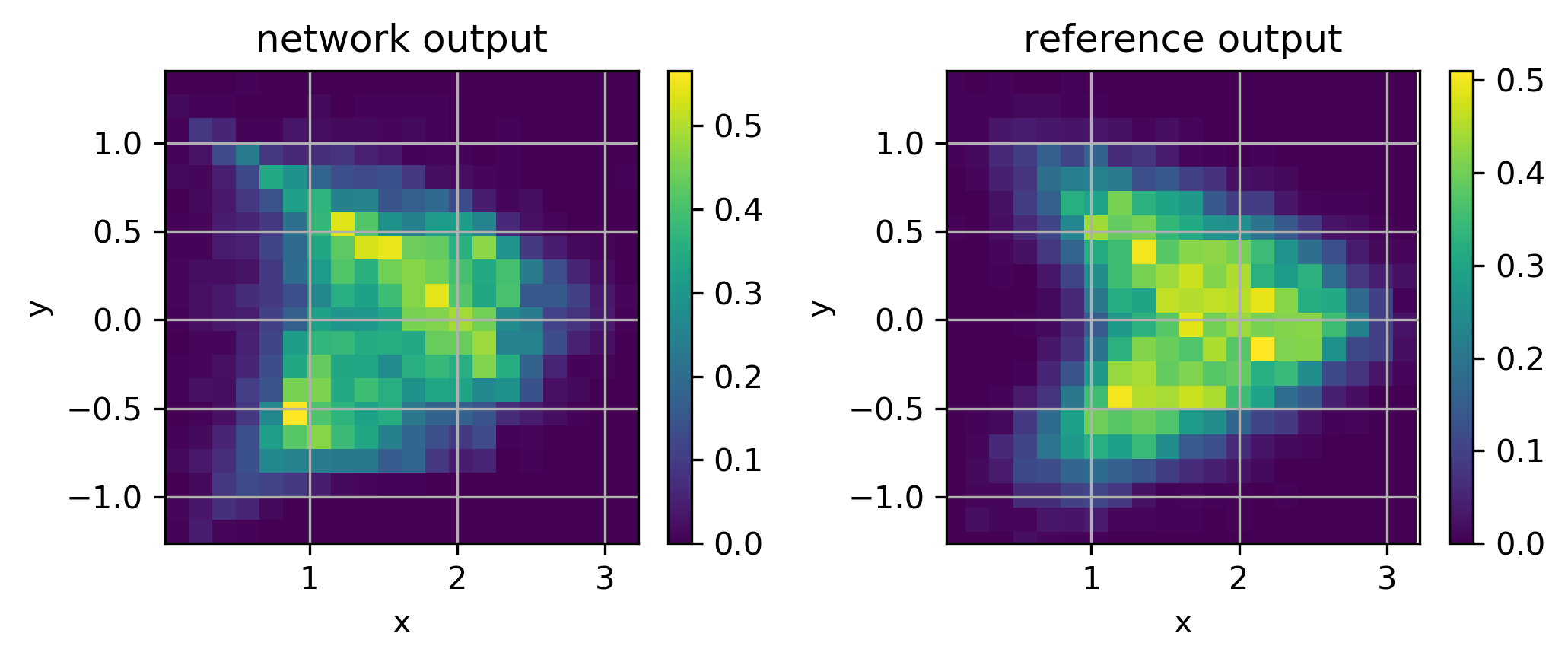}
    \caption{$t=0.02$}
    \end{subfigure}
    \begin{subfigure}{0.48\textwidth}
    \includegraphics[width=\linewidth]{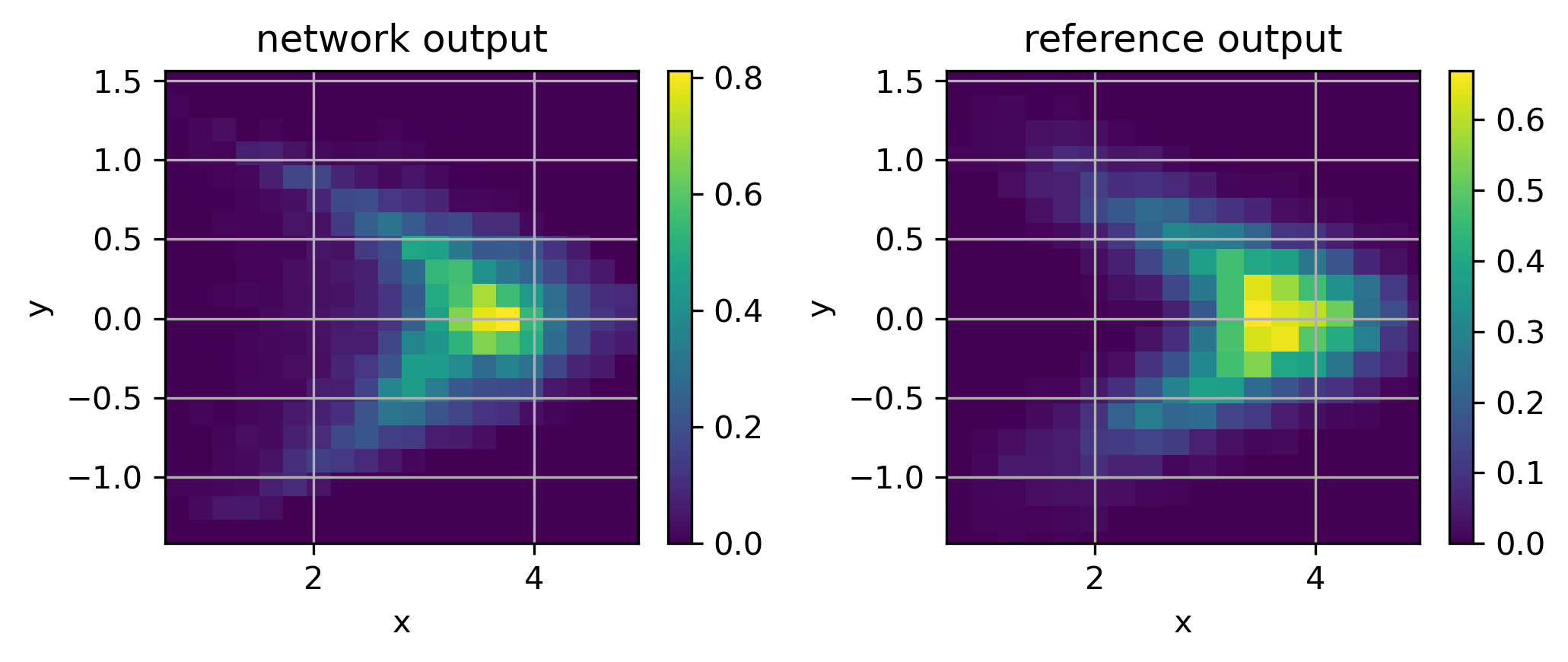}
    \caption{$t=0.04$}
    \end{subfigure}
    \begin{subfigure}{0.48\textwidth}
    \includegraphics[width=\linewidth]{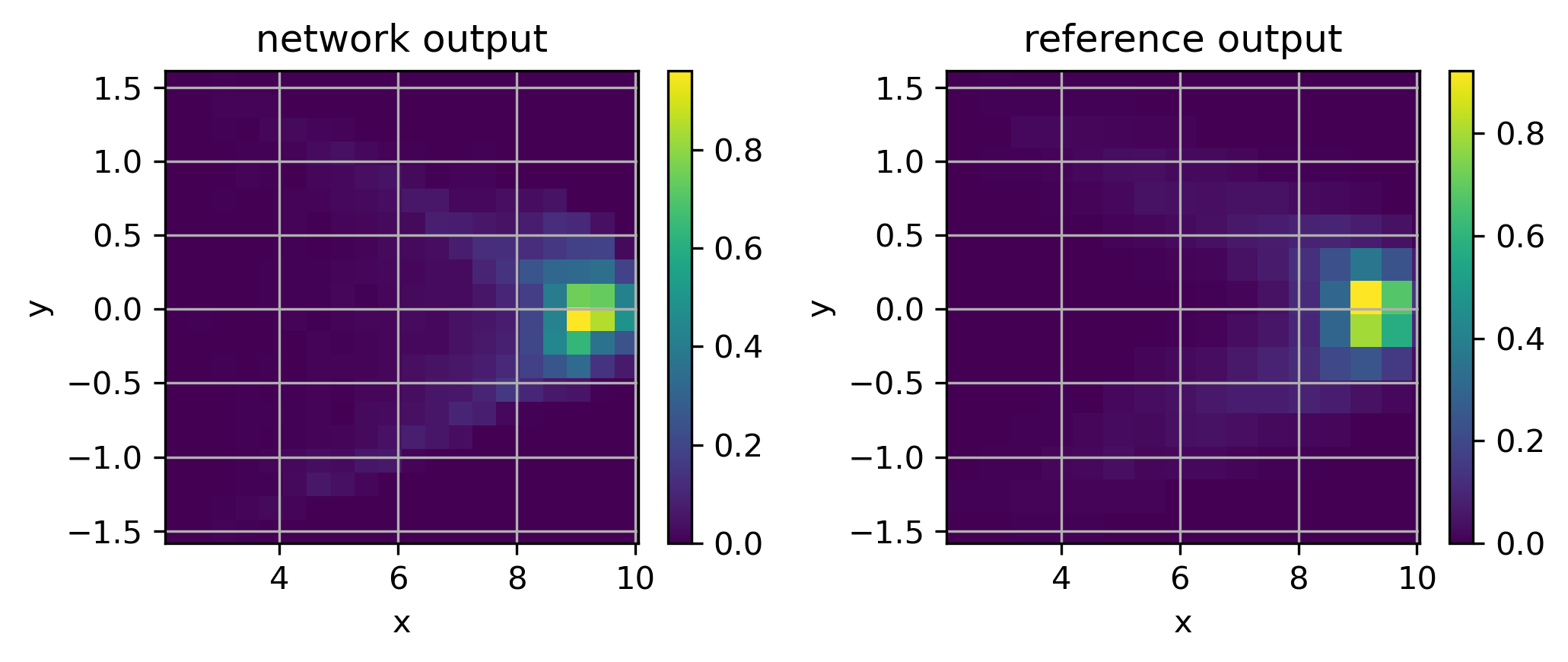}
    \caption{$t=0.1$ (largest $t$ in training set)}
    \end{subfigure}
    \begin{subfigure}{0.48\textwidth}
    \includegraphics[width=\linewidth]{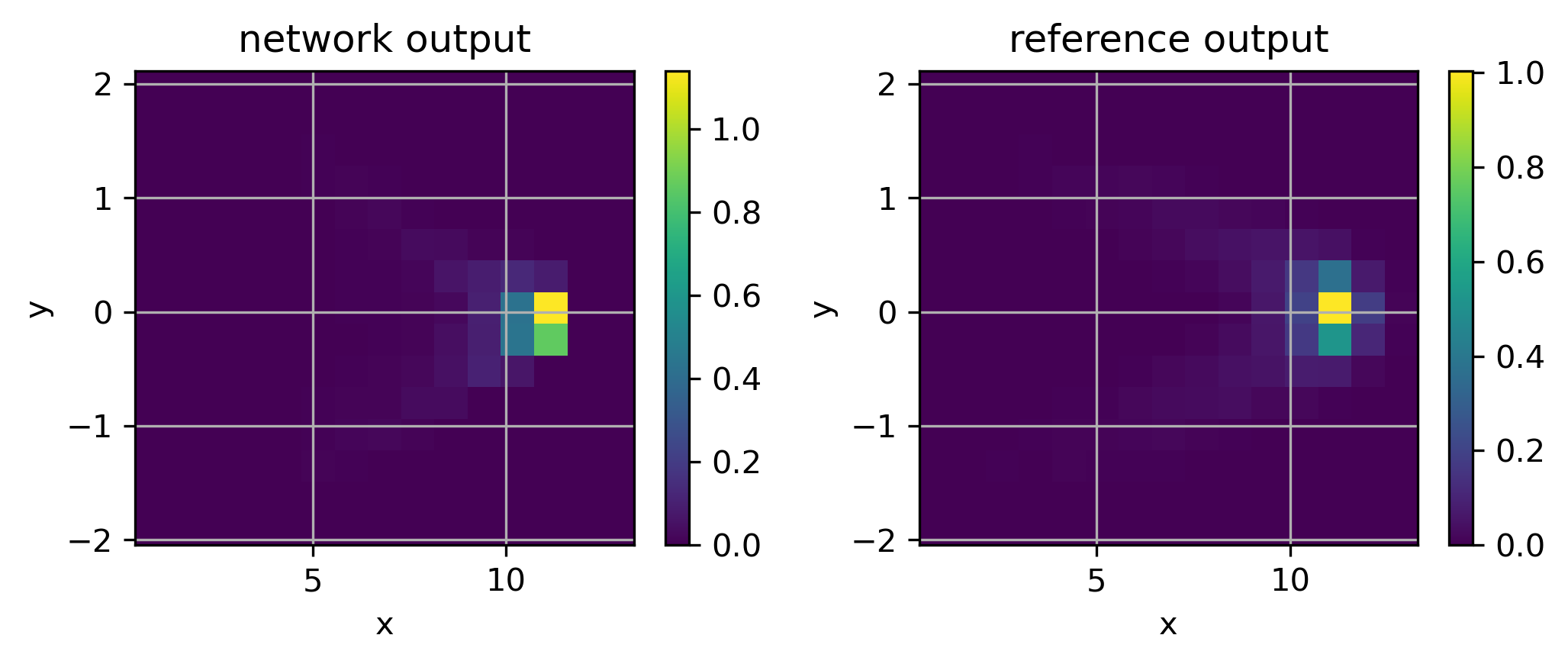}
    \caption{$t=0.14$ (extrapolation)}
    \end{subfigure}
    \caption{Learning particle aggregation at different times with fixed flow amplitude $A=100$. 
    The extrapolation performance of the network is also tested.}
    \label{fig:2DlaminerT}
\end{figure}

\paragraph{Learning the dependence on $A$} We first study the dependence of the aggregation patterns on the amplitude of the advection field while fixing the evolution time. In this example, we use the IPM to generate $J=10000$ samples of solution after $T=0.02$ with
$A=10^{0.2i}$, $i=0,....,10$.  The initial distribution of the IPM for each $A$ is a uniform distribution on the unit ball. In Fig.\ref{fig:2DlaminerC}, we can see that the distribution of the particles turns to a $V$ shape as $A$ increases. This is due to the stretching effect by the laminar flow \eqref{2Dlaminer}. 
Numerical results show that our network can learn this important feature. In addition, our network can also predict distribution when $A$ is slightly outside the range of the training set; see the case when $A=150$ in Fig.\ref{fig:2DlaminerC}(d). More precisely, Fig.\ref{fig:2DlaminerC}(c) shows that most of the outputs in the training set ($1\leq A \leq 100$) satisfy $x<3$ while at $A=150$, a reasonable proportion of particles is on the right side of $x=3$ and our network indeed predicts it.
\paragraph{Learning the dependence on evolution time}
Next, we study the dependence of the aggregation patterns on the evolution time at fixed $A=100$. To generate training data, we run the IPM with $t\in[0,0.1]$ and $J=10000$ particles, and take snapshots of the empirical distribution at $t=0,0.01,0.02,\cdots,0.1$. In Fig.\ref{fig:2DlaminerT}, we compare the output distribution 
generated by our network with the reference distribution generated by the IPM at various evolution times. We see that both IPM and DeepParticle methods reproduce the near blow-up behaviors, which are consistent with the results obtained in \cite{chemomix_yao}.
\subsection{KS Simulation and Generation in the Presence of 3D flows} In this subsection, we study the KS model with advection in three-dimensional space. There are two kinds of flows under consideration. The first flow is the 3D laminar flow which is a natural generalization of 2D laminar flow. The second one is the Kolmogorov flow which is a well-known example of chaotic flow \cite{CG95,SharpMMS_21,KLX_21}.
\paragraph{A 3D Laminar Flow}
In the first 3D example, we consider an advection field of the following form:
\begin{align}
    \textbf{v}(x,y,z)=A\, \begin{pmatrix}\exp(-y^2-z^2)\\0\\0\end{pmatrix}. \label{Laminar}
\end{align}
It describes the organism traveling along the $x$-direction while its speed depends on the radial position of $y$ and $z$ variable.  
\begin{figure}[htbp]
    \centering
    \begin{subfigure}{0.8\textwidth}
    \includegraphics[width=\textwidth]{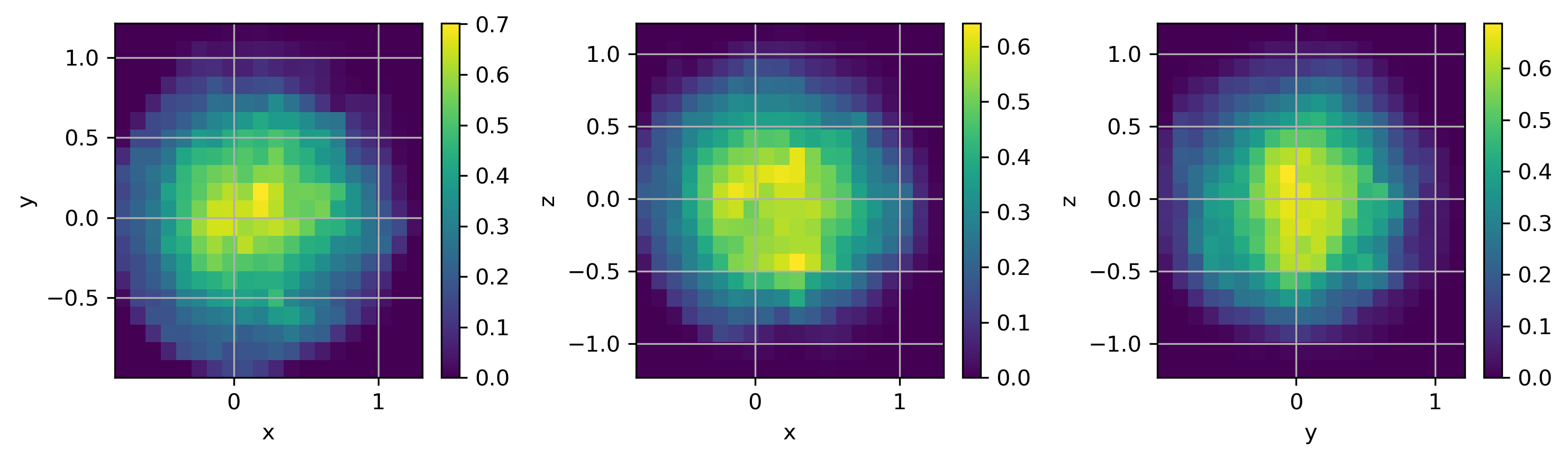}
    \caption{ $A=10$}
    \label{fig:my_label}
    \end{subfigure}
    \begin{subfigure}{0.8\textwidth}
        \includegraphics[width=\textwidth]{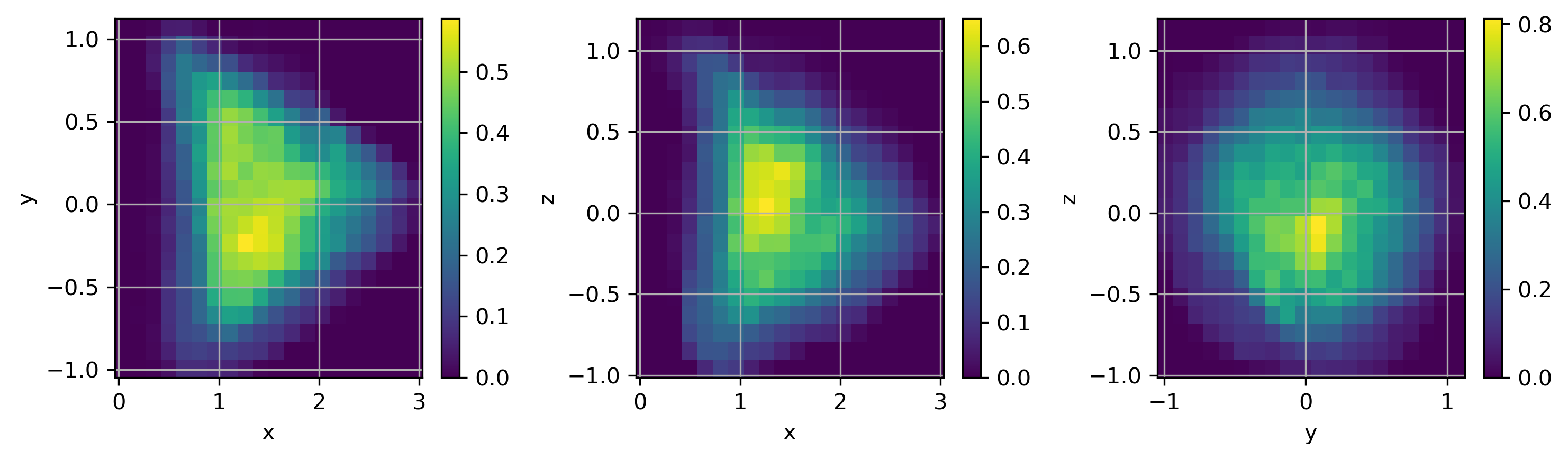}
    \caption{$A=100$}
    \label{fig:my_label}
    \end{subfigure}
    \caption{Three cross sections of generated distributions at different $A$ values in a 3D laminar flow (\ref{Laminar}).}
    \label{fig:3Dlaminar}
\end{figure}
In Fig.\ref{fig:3Dlaminar} we show the histogram of the generated distribution of our deep learning algorithms with $A=10$ and $A=100$, which reproduces the distribution of corresponding IPM simulation. The configuration of learning $A$ dependence is the same as one in 2D cases, except the input and output are now in $3$-dimension. From the numerical experiments (not shown), there is no need to increase the width of our network. By comparing (a) and (b) in Fig.\ref{fig:3Dlaminar}, we see that the distribution becomes V shape in $xy$ projection and $xz$ projection as the amplitude $A$ increases. This is due to the setting of the laminar flow. In the $yz$ projection, the distribution remains radial. 

\begin{figure}[htbp]
    \centering
     \begin{subfigure}{0.45\textwidth}
        \includegraphics[width=\textwidth]{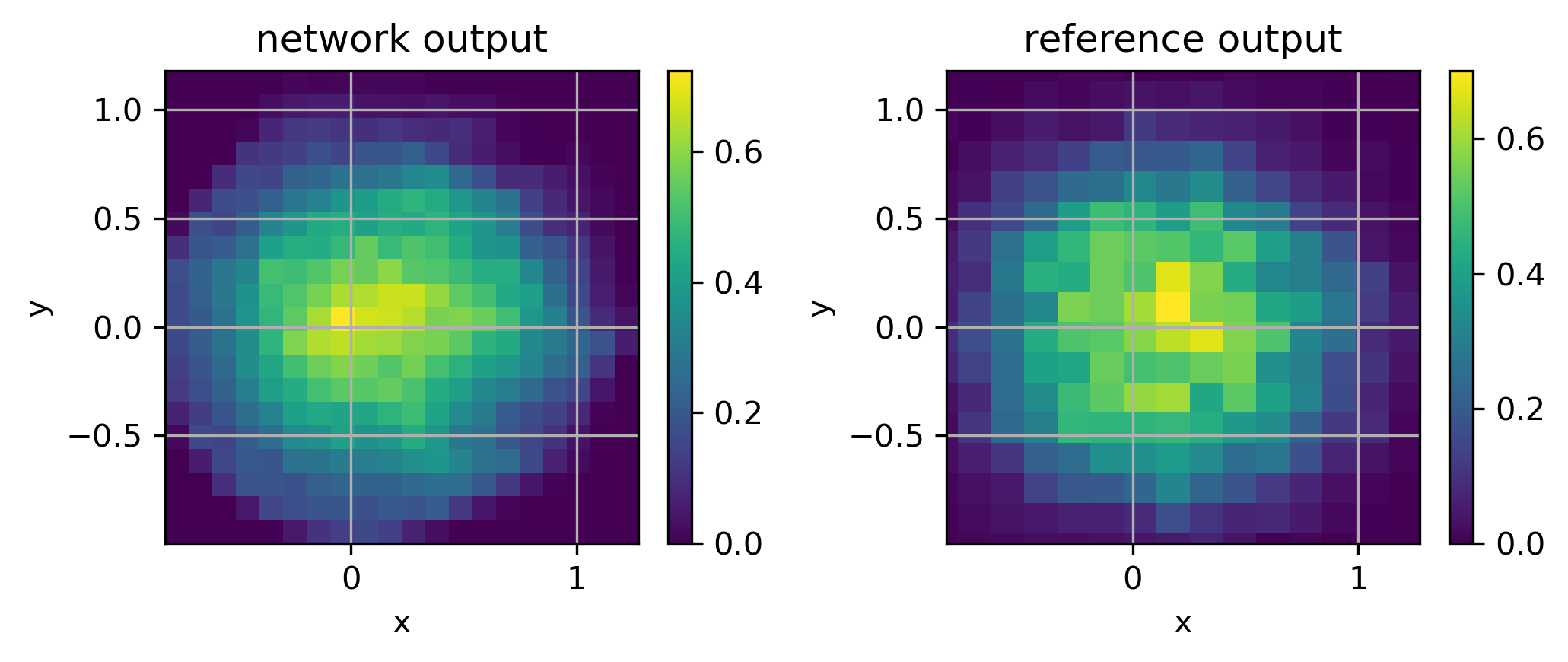}
    \caption{$A=10$}
    \label{fig:my_label}
    \end{subfigure}
      \begin{subfigure}{0.45\textwidth}
        \includegraphics[width=\textwidth]{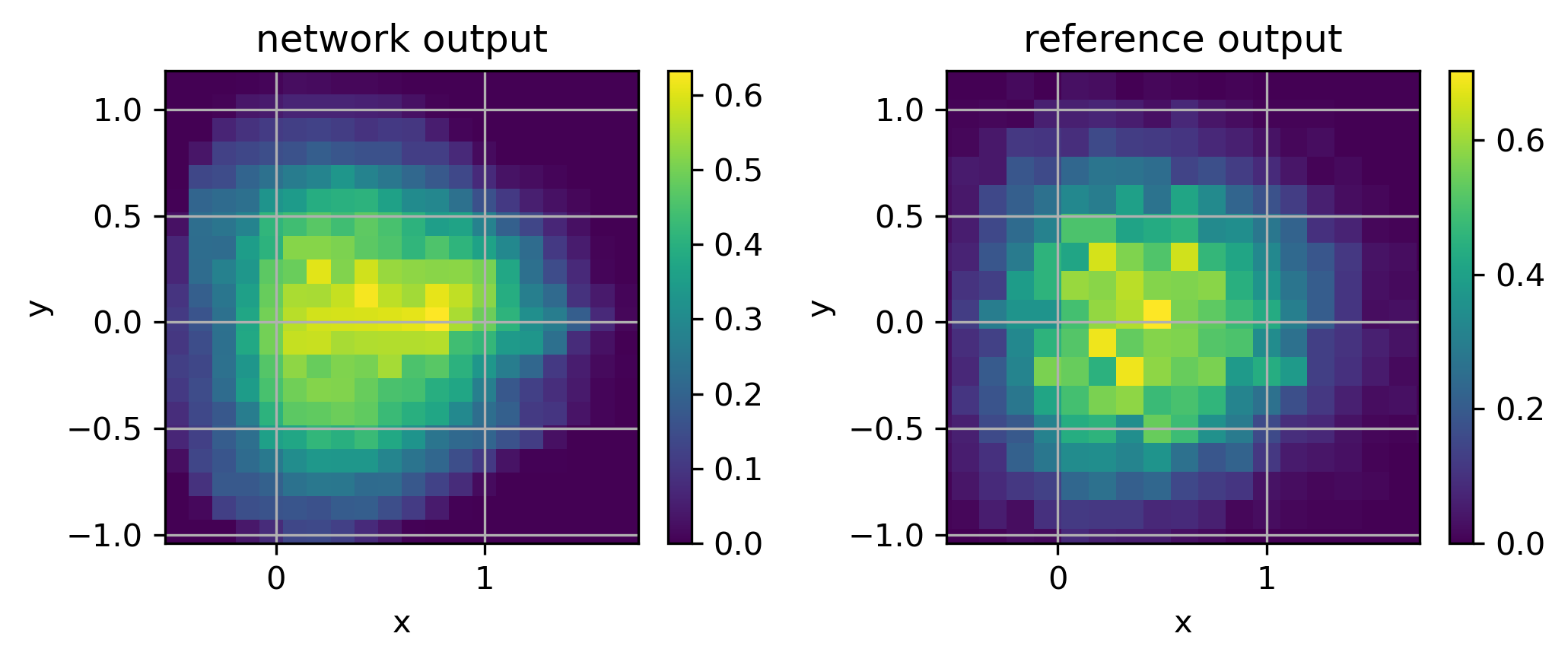}
    \caption{$A=30$}
    \label{fig:my_label}
    \end{subfigure}\\
      \begin{subfigure}{0.45\textwidth}
        \includegraphics[width=\textwidth]{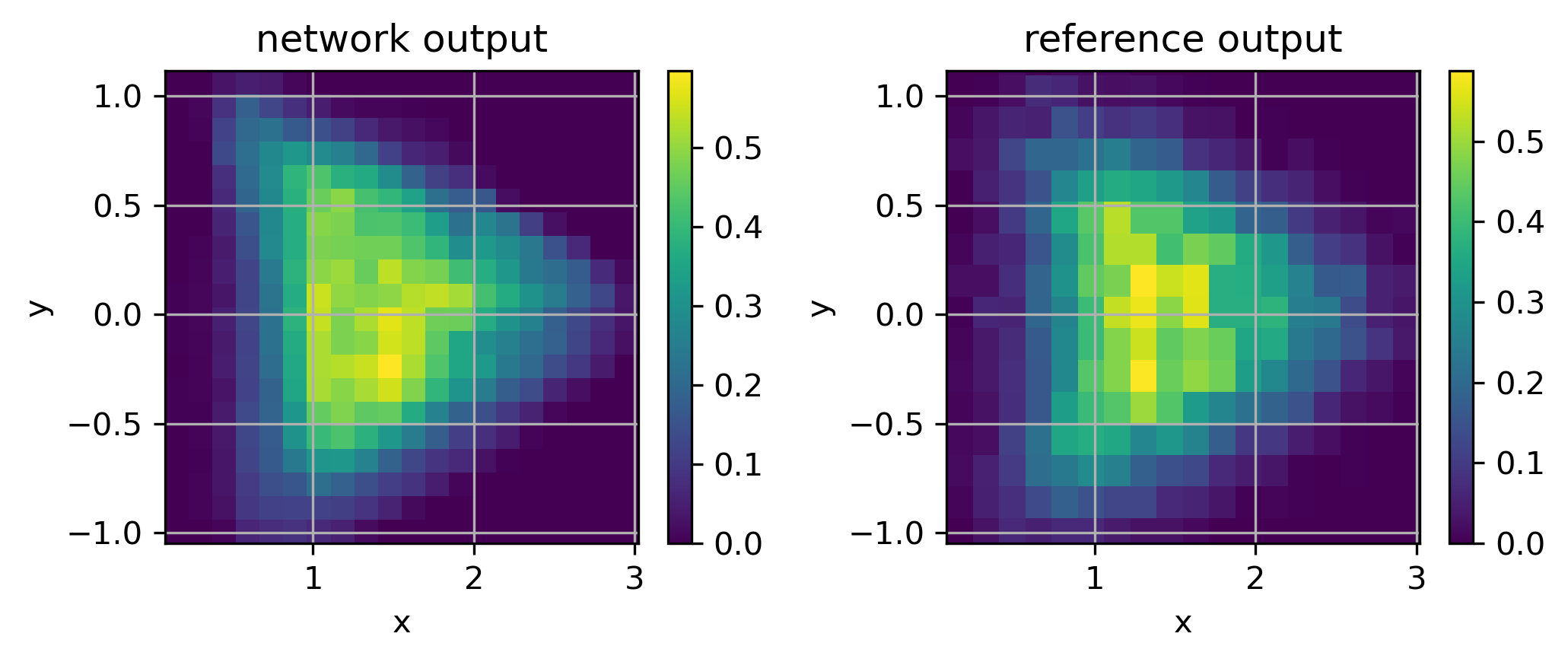}
    \caption{$A=100$}
    \label{fig:my_label}
    \end{subfigure}
      \begin{subfigure}{0.45\textwidth}
        \includegraphics[width=\textwidth]{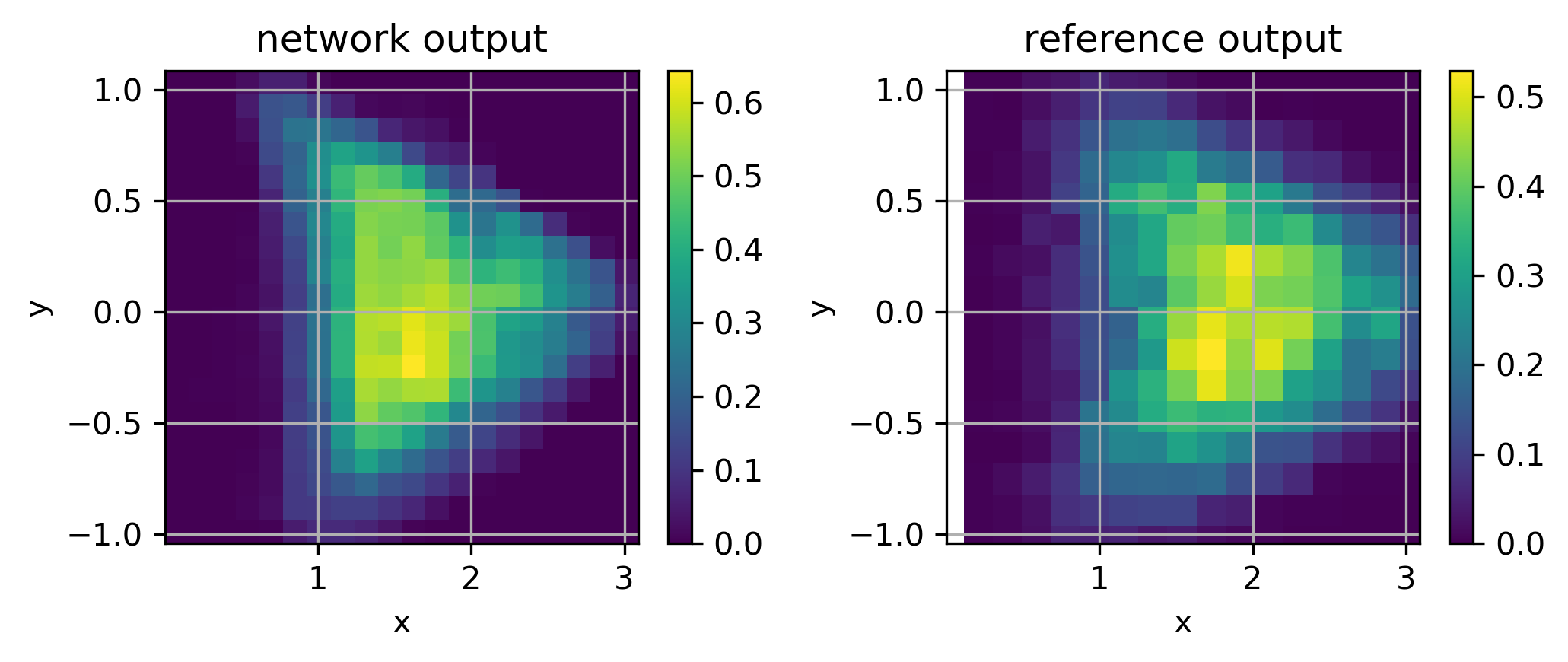}
    \caption{$A=130$ (prediction)}
    \label{fig:my_label}
    \end{subfigure}
    \caption{Network generated vs. reference distributions projected to $xy$ plane in a 3D laminar flow (\ref{Laminar}).  }
    \label{fig:3DlaminarC}
\end{figure}
In Fig.\ref{fig:3DlaminarC}, we show the $xy$ projection of the distribution with various $A$'s. It confirms that in addition to interpretation, our network is able to extrapolate (predict) the aggregation pattern associated with the amplitude $A$ that is beyond and not far from the range in the training set.
 
\begin{figure}[h]
    \centering
    \includegraphics[width=0.5\linewidth]{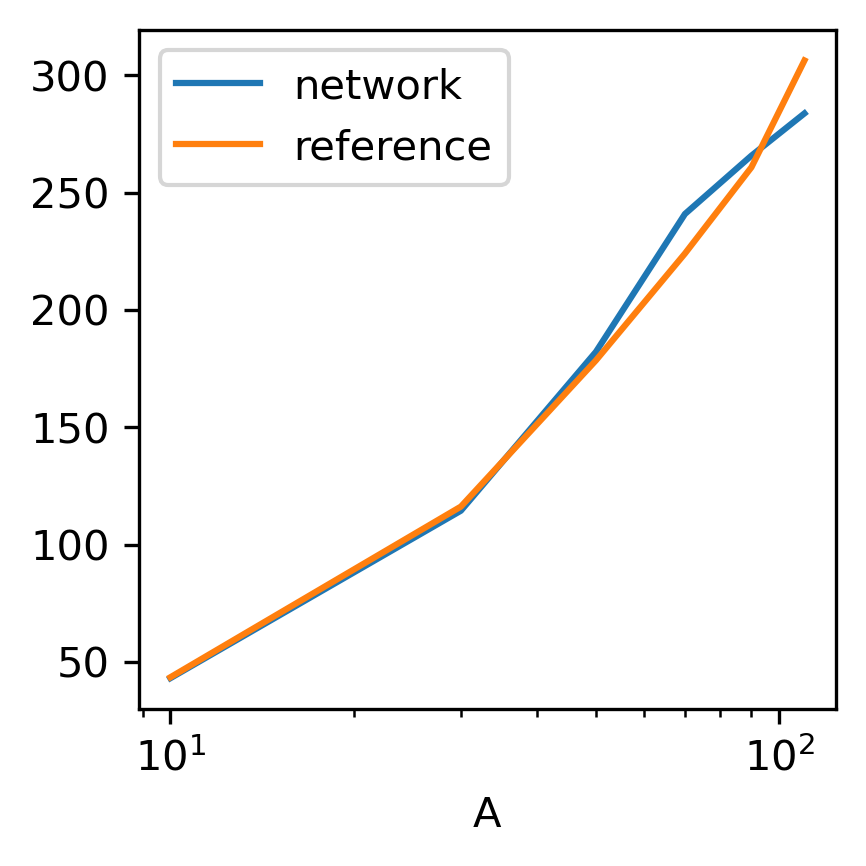}
    \caption{The second moment of distributions generated by the DeepParticle method and the reference IPM vs. the amplitude $A$ of the 3D Kolmogorov flow (\ref{Kflow}).}
    \label{fig:Kflow2nd}
\end{figure}

 {

To further investigate the generalization capabilities of our learning algorithms, we generate $J_1=10^6$ realizations using the network, denoted as $\rho_F$, and $J_2=10^4$ realizations using the IPM, denoted as $\rho_Y$. As discussed in Sec.\ref{sec:iterative_method}, direct computation of the Wasserstein distance between point cloud data $\rho_F$ and $\rho_Y$ involves linear programming with $J_1\times J_2=10^{10}$ degrees of freedom. Therefore, we consider two types of rough comparisons. 

First, we observe through the training data that the distribution of the first component increases as $A$ increases. In Fig.\ref{fig:generalization}(a), we compare the means of the first component of the generation at various $A$'s. Second, we compute an approximation of the $W^2$ distance between $\rho_F$ and $\rho_Y$. More precisely, we compute a 3D histogram of $\rho_F$ ($\rho_Y$ correspondingly) with $B^3$ uniform cells ($B=20$). Then, we approximate the distribution of $\rho_F$ by moving all points within a single cell of the histogram to the center of the cell. Finally, we compute the Wasserstein distance between the approximated distributions using the iterative method in Sec.\ref{sec:iterative_method}. See Fig.\ref{fig:generalization}(b) for the comparison of the $W^2$ distance (solid line) as well as the absolute distance of the mean (dashed line). These results show that our network can provide a reliable approximation to the reference output associated with different $A$'s within a certain range of the largest $A$ in the training set. Then, the network gradually generates larger errors if we further increase $A$.

\begin{figure}[htbp]
    \centering
     \begin{subfigure}{0.4\textwidth}
        \includegraphics[width=\textwidth]{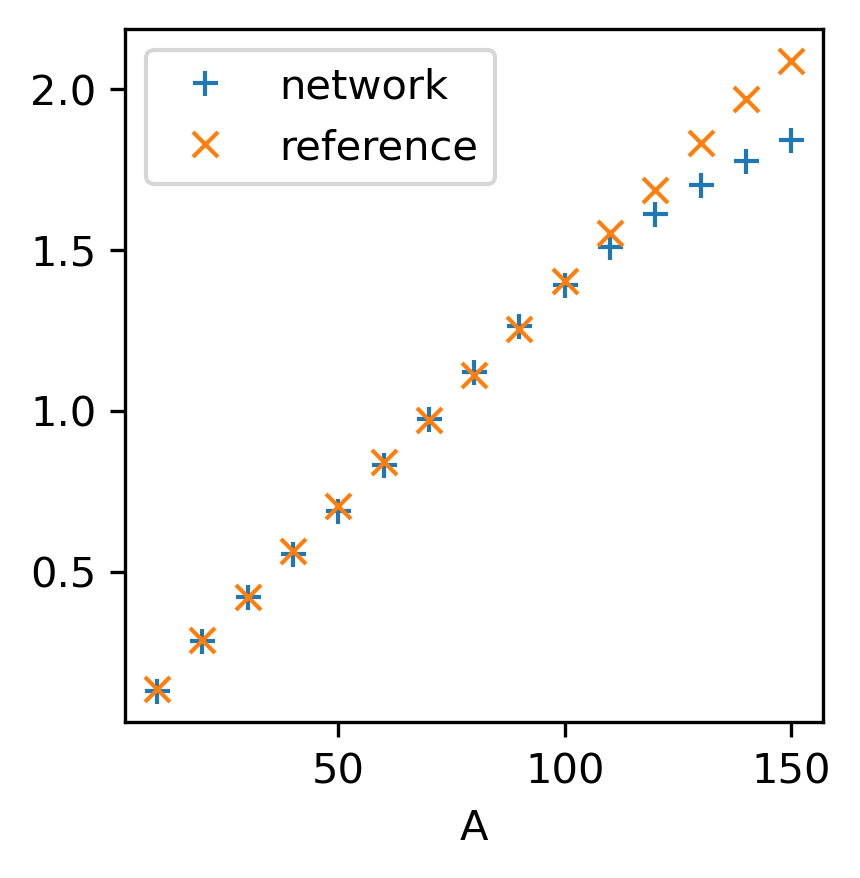}
    \caption{Mean of the first component of the generation at various $A$'s.}
    \label{fig:my_label}
    \end{subfigure}
      \begin{subfigure}{0.4\textwidth}
        \includegraphics[width=\textwidth]{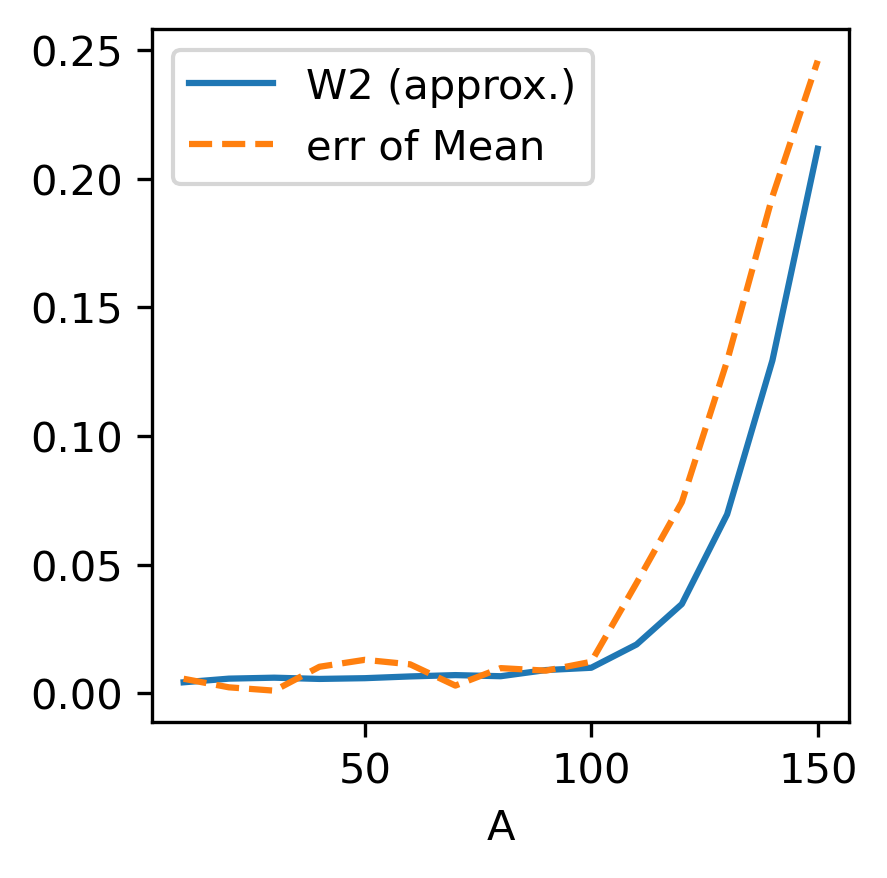}
    \caption{Comparison between the approximated $W^2$ distance and the sample mean.}
    \label{fig:my_label}
    \end{subfigure}
    \caption{Comparison of the generation at various $A$'s in a 3D laminar flow (\ref{Laminar}). }
    \label{fig:generalization}
    \end{figure}

}
 \paragraph{The 3D Kolmogorov Flow}
In the second example, we investigate the case when the organism travels and aggregates in chaotic streamlines given by the Kolmogorov flow \cite{CG95,SharpMMS_21,KLX_21}:

\begin{align}
    \textbf{v}(x,y,z)=A\, \begin{pmatrix}\sin(2\pi z)\\\sin(2\pi x)\\\sin(2\pi y)\end{pmatrix}. \label{Kflow}
\end{align}
\begin{figure}[htbp]
    \centering
    \includegraphics[width=0.9\textwidth]{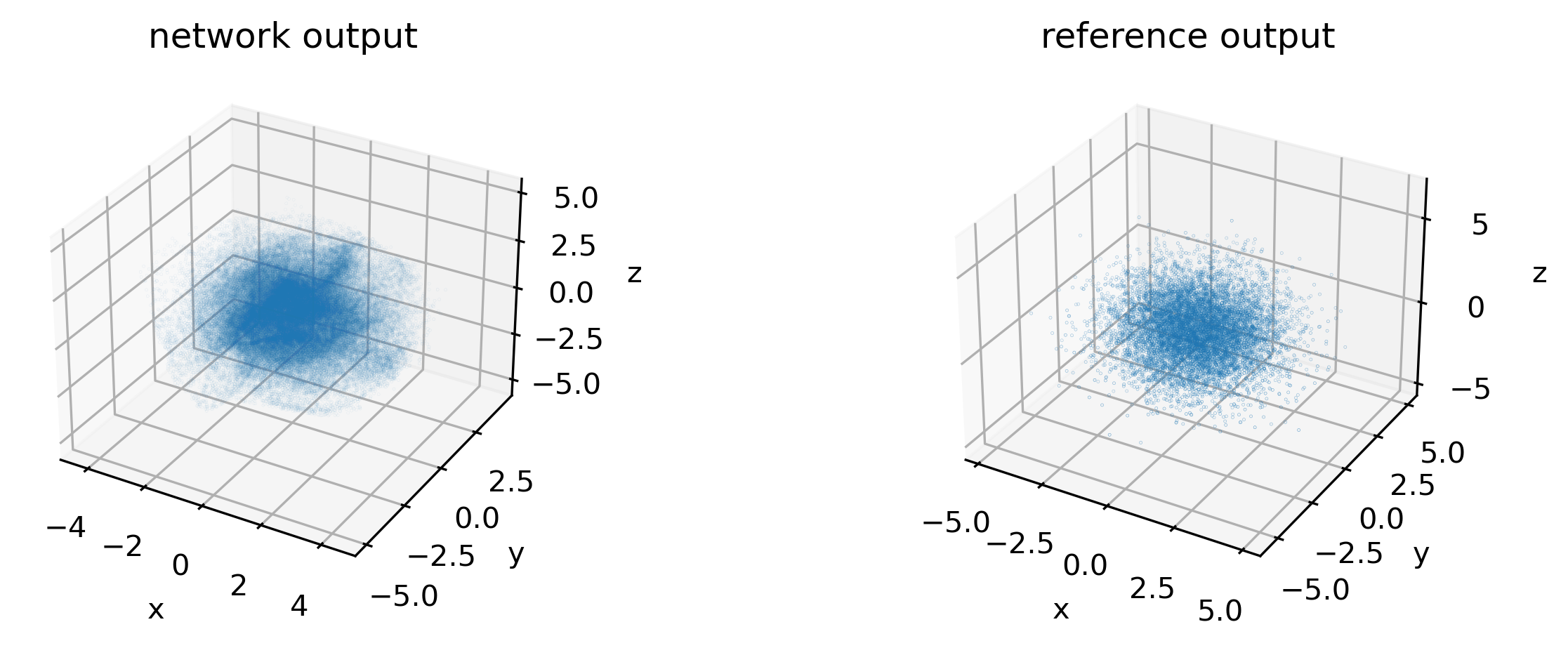}
    \caption{Network output ($N=1M$) vs. training data ($N=10K$) for the Kolmogorov flow (\ref{Kflow}) with $A=100$, where the evolution time $t=0.1$.}
        \label{fig:Kflow}
\end{figure}
\begin{figure}[htbp]
    \centering
     \begin{subfigure}{0.49\textwidth}
        \includegraphics[width=\textwidth]{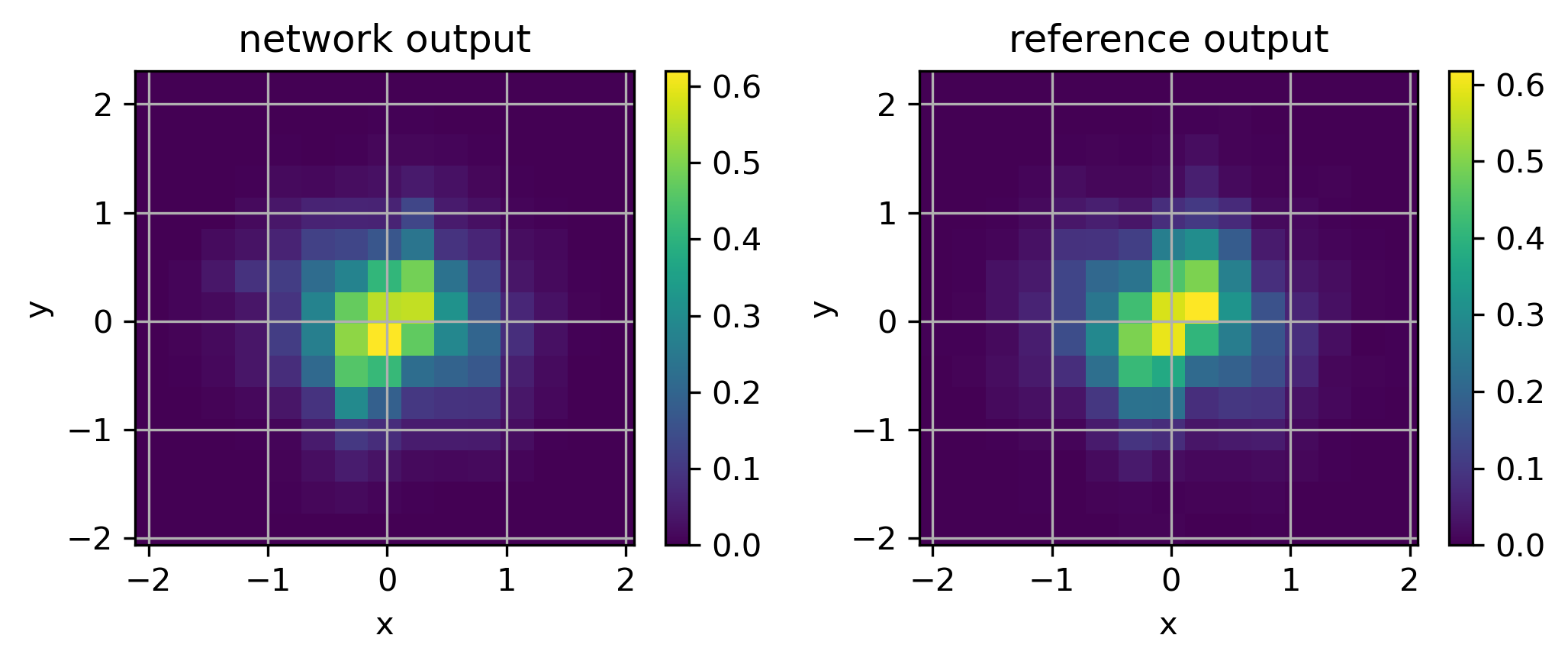}
    \caption{$A=10$}
    \label{fig:my_label}
    \end{subfigure}
      \begin{subfigure}{0.49\textwidth}
        \includegraphics[width=\textwidth]{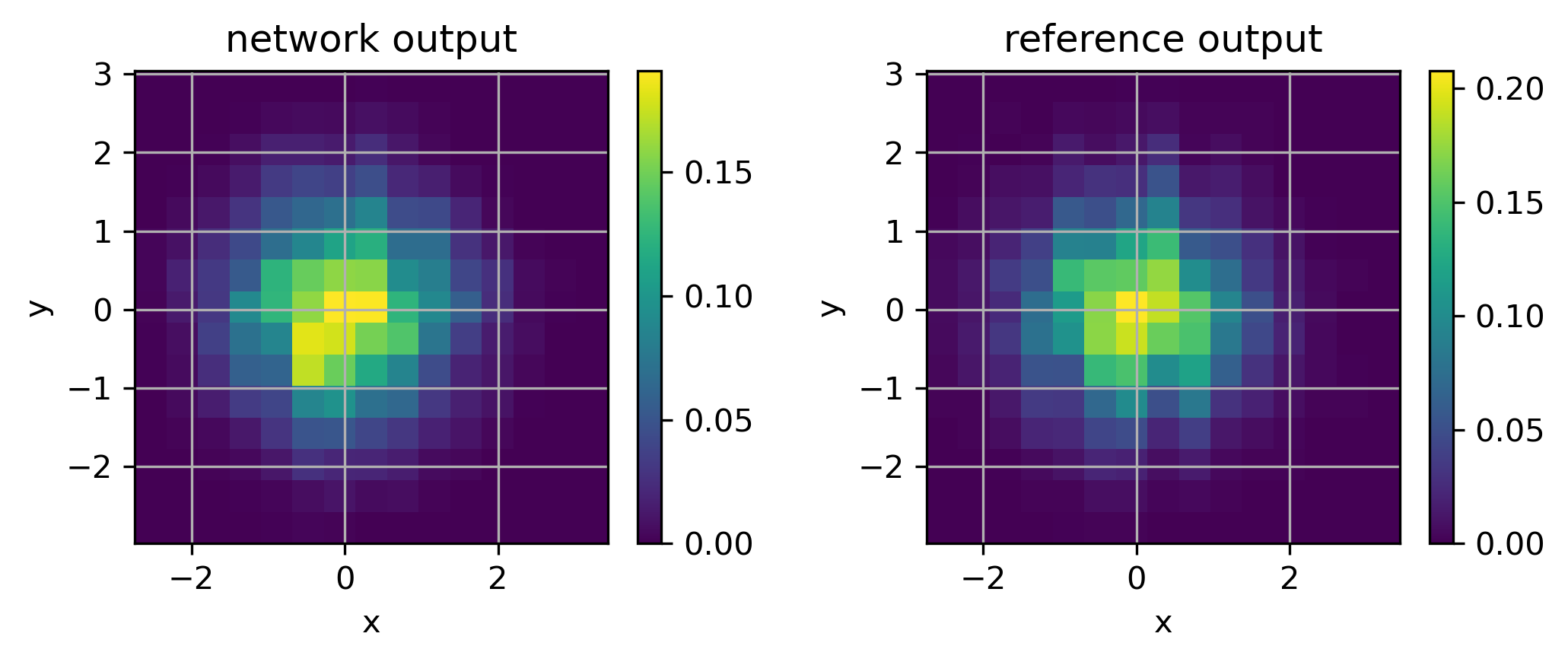}
    \caption{$A=30$}
    \label{fig:my_label}
    \end{subfigure}\\
      \begin{subfigure}{0.49\textwidth}
        \includegraphics[width=\textwidth]{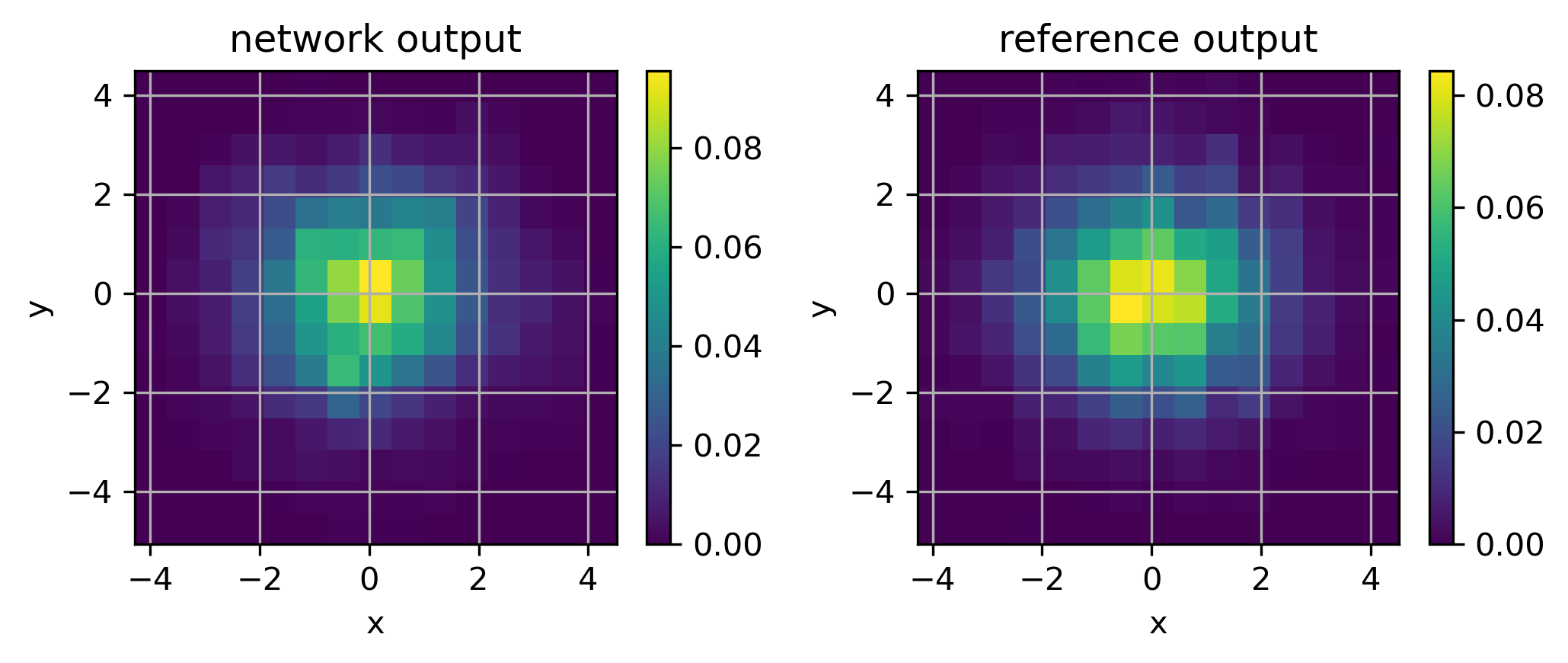}
    \caption{$A=100$}
    \label{fig:my_label}
    \end{subfigure}
      \begin{subfigure}{0.49\textwidth}
        \includegraphics[width=\textwidth]{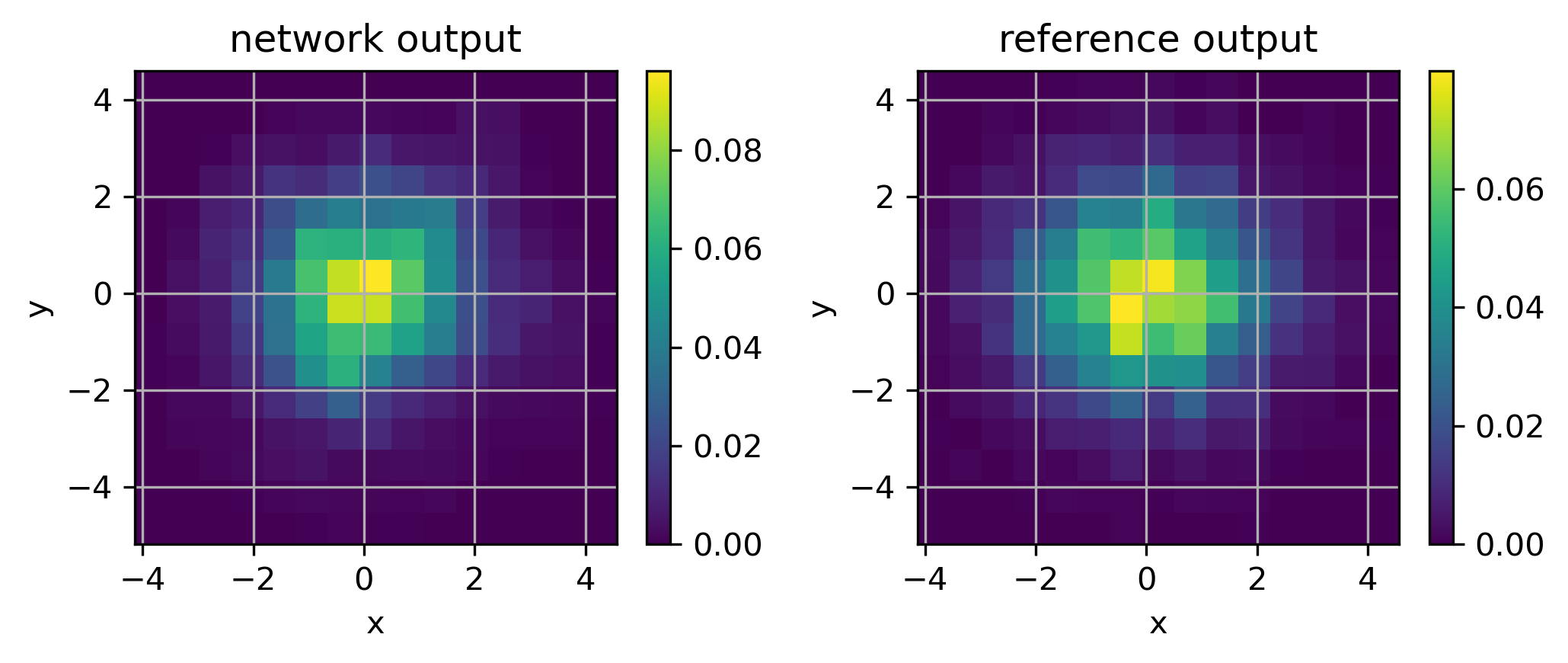}
    \caption{$A=110$ (prediction)}
    \label{fig:my_label}
    \end{subfigure}
    \caption{Network generated vs. reference distributions projected to $xy$ plane in the 3D Kolmogorov flow (\ref{Kflow}).}
    \label{fig:KflowC}
\end{figure}
In Fig.\ref{fig:Kflow}, we compare the distributions generated by the network method and the reference solver (i.e., the IPM) when $A=100$ and $t=0.1$. { This demonstrates that our network method, after learning from discrete empirical data, is capable of generating continuous distributions.} And in Fig.\ref{fig:KflowC}, we compare the distributions associated with various amplitude $A$'s when projected to $xy$ plane. The distributions are in general a radial distribution and the radius of the distribution increases when $A$ increases (see also the second moment plot in Fig.\ref{fig:Kflow2nd}). The phenomenon differs from the one in laminar flow. This may result from the mixing mechanism of the chaotic flow that spreads and acts against the chemotaxis aggregation. 

Finally, we summarize the performance of our algorithms measured by the Wasserstein distance in the validation and prediction data set in Table \ref{tab:performance} as follows. These results show that the DeepParticle method 
is an efficient method for learning and generating aggregation patterns in 2D and 3D KS chemotaxis systems.

  \begin{table}[h]
  \centering 
  \begin{threeparttable}
    \begin{tabular}{ccrcr}
    \toprule
Model & Validation Config. &  $W^2$ & Prediction Config. & $W^2$ \\ 
\midrule
2D Laminar $A=0$ & $t=0.05$ & 0.0086 & $t=0.12$ & 0.0120 \\
2D Laminar $A=100$ & $t=0.05$ & 0.0116 & $t=0.12$ & 0.0190 \\
2D Laminar $t=0.02$ & $A=50$ & 0.0041 & $A=130$ & 0.0311 \\
3D Laminar $t=0.02$ & $A=50$ & 0.0217 & $A=130$ & 0.0946 \\
3D K flow $t=0.02$ & $A=50$ & 0.0410 & $A=110$ & 0.0619 \\
  \bottomrule
    \end{tabular}
\end{threeparttable}
 \caption{Wasserstein distance between the network output and reference output in various cases.}
 \label{tab:performance}
\end{table}
\section{Conclusions and Future Works}
 
We proposed a regularized interacting particle method (IPM) to compute aggregation patterns and near singular solutions in multi-dimensional KS systems.  We then studied a DeepParticle method to learn and generate solutions for KS systems with dependence on physical parameters (e.g. the flow amplitude in the advection-dominated regime and the evolution time) by minimizing the 2-Wasserstein distance between the source and target distributions. During the training stage, we seek a mapping in the form of a deep neural network from source to target distributions and update network parameters based on a discretized 2-Wasserstein distance defined on finite distribution samples. Our method is general in the sense that we do not require target distributions to be in closed form and the generation map to be invertible. Our method is fully data-driven and applicable to the fast generation of distributions for more general KS systems with physical parameter dependency. Our iterative divide-and-conquer algorithm reduces considerably the computational cost of finding the optimal transition matrix in the Wasserstein distance. We carried out numerical experiments to demonstrate the performance of our method 
for learning and generating aggregation patterns in 2D and 3D KS chemotaxis systems 
without and with laminar and chaotic advection. 

In future work, we plan to study the DeepParticle method to learn and generate pattern-forming solutions of parabolic type KS systems ($\epsilon > 0$ in (\ref{KS1})) among other KS like (e.g. chemotaxis-haptotaxis) systems for modeling and predicting cancer cell evolution \cite{chaplain2006}.

\section*{CRediT authorship contribution statement}
\noindent
\textbf{Zhongjian Wang}: Conceptualization, Programming, Methodology, Writing-Original draft preparation and Editing. \textbf{Jack Xin}: Conceptualization, Methodology, Writing-Reviewing, and Editing. \textbf{Zhiwen Zhang}: Conceptualization, Methodology, Writing-Reviewing, and Editing.

\section*{Declaration of competing interest}
\noindent
The authors declare that they have no known competing financial interests or personal relationships that could have appeared to influence the work reported in this paper.

\section*{Acknowledgements}
\noindent
The research of ZW is partially supported by NTU SUG-023162-00001. The research of JX is partially supported by NSF grants DMS-1952644, and DMS-2309520. The research of ZZ is supported by the Hong Kong RGC grant (Projects 17300318 and 17307921), the National Natural Science Foundation of China  (Project 12171406), Seed Funding Programme for Basic Research (HKU), the outstanding young researcher award of HKU (2020-21), and Seed Funding for Strategic Interdisciplinary Research Scheme 2021/22 (HKU). The authors would like to thank Prof. John Lowengrub at UC Irvine and Prof. Philip Maini at Oxford University for helpful discussions of chemotaxis, cell growth, and cancer modeling.

\bibliographystyle{plain}
\bibliography{ZWpaper_DeepLearningPDEDiscCoef} 

\begin{thebibliography}{10}

\bibitem{arjovsky2017wasserstein}
M.~Arjovsky, S.~Chintala, and L.~Bottou.
\newblock Wasserstein generative adversarial networks.
\newblock In {\em International conference on machine learning}, pages
  214--223. PMLR, 2017.

\bibitem{Osher_21}
M.~Burger, L.~Ruthotto, and S.~Osher.
\newblock Connections between deep learning and partial differential equations.
\newblock {\em European Journal of Applied Mathematics}, 32(3):395--396, 2021.

\bibitem{caflisch1989singular}
R.~Caflisch and O.~Orellana.
\newblock Singular solutions and ill-posedness for the evolution of vortex
  sheets.
\newblock {\em SIAM Journal on Mathematical Analysis}, 20(2):293--307, 1989.

\bibitem{blob_diff_2019}
J.~Carrillo, K.~Craig, and F.~Patacchini.
\newblock A blob method for diffusion.
\newblock {\em Calculus of Variations}, 58(53), 2019.

\bibitem{carrillo2019hybrid}
J.~Carrillo, N.~Kolbe, and M.~Luk{\'a}{\v{c}}ov{\'a}.
\newblock A hybrid mass transport finite element method for {K}eller--{S}egel
  type systems.
\newblock {\em Journal of Scientific Computing}, 80(3):1777--1804, 2019.

\bibitem{chaplain2006}
M.~Chaplain and G.~Lolas.
\newblock Mathematical modelling of cancer invasion of tissue: Dynamic
  heterogeneity.
\newblock {\em Networks and Heterogeneous Media}, 1(3):399–439, 2006.

\bibitem{CG95}
S.~Childress and A.~Gilbert.
\newblock {\em Stretch, Twist, Fold: The Fast Dynamo}.
\newblock Lecture Notes in Physics Monographs, No. 37, Springer, 1995.

\bibitem{chorin1973discretization}
A.~Chorin and P.~Bernard.
\newblock Discretization of a vortex sheet, with an example of roll-up.
\newblock {\em Journal of Computational Physics}, 13(3):423--429, 1973.

\bibitem{blob_aggregation_ck2016}
K.~Craig and A.~Bertozzi.
\newblock A blob method for the aggregation equation.
\newblock {\em Math. Comp.}, 85(300):1681--1717, 2016.

\bibitem{fatkullin2012study}
I.~Fatkullin.
\newblock A study of blow-ups in the {K}eller--{S}egel model of chemotaxis.
\newblock {\em Nonlinearity}, 26(1):81, 2012.

\bibitem{havskovec2009stochastic}
J.~Ha{\v{s}}kovec and C.~Schmeiser.
\newblock Stochastic particle approximation for measure valued solutions of the
  2{D} {K}eller-{S}egel system.
\newblock {\em Journal of Statistical Physics}, 135(1):133--151, 2009.

\bibitem{chemomix_22}
S.~He, E.~Tadmor, and A.~Zlato\v{s}.
\newblock On the fast spreading scenario.
\newblock {\em Comm. Amer. Math. Soc.}, 2:149--171, 2022.

\bibitem{hou2009stabconv}
T.~Hou and Z.~Lei.
\newblock On the stabilizing effect of convection in 3{D} incompressible flow.
\newblock {\em Commun. Pure Appl. Math.}, 62(4):501--564, 2009.

\bibitem{hou2006dynamic}
T.~Hou and R.~Li.
\newblock Dynamic depletion of vortex stretching and non-blowup of the 3{D}
  incompressible {E}uler equations.
\newblock {\em Journal of Nonlinear Science}, 16(6):639--664, 2006.

\bibitem{chemomix_21}
G.~Iyer, X.~Xu, and A.~Zlato\v{s}.
\newblock Convection-induced singularity suppression in the {K}eller-{S}egel
  and other non-linear {PDE}s.
\newblock {\em Trans. Amer. Math. Soc.}, 374:6039--6058, 2021.

\bibitem{KLX_21}
C.~Kao, Y-Y Liu, and J.~Xin.
\newblock A {Semi-Lagrangian} {Computation of Front Speeds of G-equation in ABC
  and Kolmogorov Flows with Estimation via Ballistic Orbits}.
\newblock {\em Multiscale Modeling and Simulation}, 20(1):107--117, 2022.

\bibitem{keller1970initiation}
E.~Keller and L.~Segel.
\newblock Initiation of slime mold aggregation viewed as an instability.
\newblock {\em Journal of theoretical biology}, 26(3):399--415, 1970.

\bibitem{chemomix_yao}
S.~Khan, J.~Johnson, E.~Cartee, and Y.~Yao.
\newblock Global regularity of chemotaxis equations with advection.
\newblock {\em Involve}, 9(1):119–131, 2016.

\bibitem{chemomix_12}
A.~Kiselev and L.~Ryzhik.
\newblock Biomixing by chemotaxis and enhancement of biological reactions.
\newblock {\em Communications in PDE}, 37:298–318, 2012.

\bibitem{chemomix_16}
A.~Kiselev and X.~Xu.
\newblock Suppression of chemotactic explosion by mixing.
\newblock {\em Arch. Ration. Mech. Anal.}, 222(2):1077–1112, 2016.

\bibitem{krasny1986desingularization}
R.~Krasny.
\newblock Desingularization of periodic vortex sheet roll-up.
\newblock {\em Journal of Computational Physics}, 65(2):292--313, 1986.

\bibitem{krasny1991vortex}
R.~Krasny.
\newblock Vortex sheet computations: roll-up, wakes, separation.
\newblock {\em Lectures in Applied Mathematics}, 28(1):385--401, 1991.

\bibitem{liu1995convergence}
J.~Liu and Z.~Xin.
\newblock Convergence of vortex methods for weak solutions to the 2{D} {E}uler
  equations with vortex sheet data.
\newblock {\em Communications on Pure and Applied Math}, 48(6):611--628, 1995.

\bibitem{rpbm_2017}
J.~Liu and R.~Yang.
\newblock A random particle blob method for the {K}eller-{S}egel equation and
  convergence analysis.
\newblock {\em Mathematics of Computation}, 86(304):725--745, 2017.

\bibitem{IPM_2020}
J.~Lyu, Z.~Wang, J.~Xin, and Z.~Zhang.
\newblock Convergence analysis of stochastic structure-preserving schemes for
  computing effective diffusivity in random flows.
\newblock {\em SIAM J. Numer. Anal.}, 58(5):3040--3067, 2020.

\bibitem{IPM_2021}
J.~Lyu, Z.~Wang, J.~Xin, and Z.~Zhang.
\newblock A convergent interacting particle method and computation of {KPP}
  front speeds in chaotic flows.
\newblock {\em SIAM J. Numer. Anal.}, 60(3):1136--1167, 2022.

\bibitem{Oth_97}
H.~Othmer and A.~Stevens.
\newblock Aggregation, blowup, and collapse: the {ABC}’s of taxis in
  reinforced random walks.
\newblock {\em SIAM J. Appl. Math}, 57:1044–1081, 1997.

\bibitem{Patlak_53}
C.~Patlak.
\newblock Random walk with persistence and external bias.
\newblock {\em Bull. Math. Biol.}, 15:311–338, 1953.

\bibitem{perthame2004pde}
B.~Perthame.
\newblock {PDE} models for chemotactic movements: parabolic, hyperbolic and
  kinetic.
\newblock {\em Applications of Mathematics}, 49(6):539--564, 2004.

\bibitem{schrijver2003combinatorial}
A.~Schrijver.
\newblock {\em Combinatorial optimization: polyhedra and efficiency},
  volume~24.
\newblock Springer Science \& Business Media, 2003.

\bibitem{shen2020unconditionally}
J.~Shen and J.~Xu.
\newblock Unconditionally bound preserving and energy dissipative schemes for a
  class of {K}eller--{S}egel equations.
\newblock {\em SIAM Journal on Numerical Analysis}, 58(3):1674--1695, 2020.

\bibitem{sinkhorn1964relationship}
R.~Sinkhorn.
\newblock A relationship between arbitrary positive matrices and doubly
  stochastic matrices.
\newblock {\em The Annals of Mathematical Statistics}, 35(2):876--879, 1964.

\bibitem{villani2021topics}
C.~Villani.
\newblock {\em Topics in optimal transportation}, volume~58.
\newblock American Math. Soc., 2021.

\bibitem{WangXinZhang:18}
Z.~Wang, J.~Xin, and Z.~Zhang.
\newblock Computing effective diffusivity of chaotic and stochastic flows using
  structure-preserving schemes.
\newblock {\em SIAM J. Numer. Anal}, 56(4):2322--2344, 2018.

\bibitem{SharpMMS_21}
Z.~Wang, J.~Xin, and Z.~Zhang.
\newblock Sharp uniform in time error estimate on a stochastic
  structure-preserving {L}agrangian method and computation of effective
  diffusivity in 3{D} chaotic flows.
\newblock {\em Multiscale Modeling and Simulation}, 93(3):1167--1189, 2021.

\bibitem{TDChaoticFlow_22}
Z.~Wang, J.~Xin, and Z.~Zhang.
\newblock Computing effective diffusivities of 3{D} time-dependent chaotic
  flows with a convergent {L}agrangian numerical method.
\newblock {\em ESAIM: Mathematical Modeling and Numerical Analysis},
  56:1521--1544, 2022.

\bibitem{DP_22}
Z.~Wang, J.~Xin, and Z.~Zhang.
\newblock Deep{P}article: learning invariant measure by a deep neural network
  minimizing {W}asserstein distance on data generated by an interacting
  particle method.
\newblock {\em J. Computational Physics}, 464:111309, 2022.

\bibitem{wright1997primal}
S.~Wright.
\newblock {\em Primal-dual interior-point methods}.
\newblock SIAM Publications, Philadelphia, 1997.

\end{thebibliography}
\end{document}